\input ./style/arxiv-general.cfg
\documentclass[aoas,MSNbibl,nameyear,dvips]{arximspdf}
\makeatletter
   \@ifpackageloaded{graphicx}{}{\usepackage{graphicx}}
\makeatother
\usepackage{algorithm,multirow}

\doi{10.1214/15-AOAS859}
\volume{9}
\issue{4}
\pubyear{2015}
\firstpage{1823}
\lastpage{1863}
\docsubty{FLA}

\makeatletter
\newcommand{\lleft}{\left}
\newcommand{\rrvert}{\vert}
\newcommand{\rright}{\right}
\newcommand{\rrVert}{\Vert}
\newcommand{\llvert}{\vert}
\newcommand{\llVert}{\Vert}
\renewcommand{\mid}{|}
\newcommand{\eqref}[1]{(\ref{#1})}
\newtheorem{proposition}{Proposition}
\newtheorem{theorem}{Theorem}
\newtheorem{lemma}{Lemma}
\makeatother

\begin{document}
\begin{frontmatter}

\title{The latent state hazard model, with application to~wind turbine reliability}
\runtitle{The latent state hazard model}

\begin{aug}
\author[A]{\fnms{Ramin}~\snm{Moghaddass}\corref{}\ead[label=e1]{raminm@mit.edu}}
\and
\author[B]{\fnms{Cynthia}~\snm{Rudin}\ead[label=e2]{cynthia@mit.edu}}
\runauthor{R. Moghaddass and C. Rudin}
\affiliation{Massachusetts Institute of Technology}
\address[A]{Sloan School of Management\\
Massachusetts Institute of Technology\\
Cambridge, Massachusetts 02142\\
USA\\
\printead{e1}}
\address[B]{Computer Science and\\
\quad Artificial Intelligence Laboratory\\
Massachusetts Institute of Technology\\
Cambridge, Massachusetts 02142\\
USA\\
\printead{e2}}
\end{aug}

%
\received{\smonth{9} \syear{2014}}
%
\revised{\smonth{7} \syear{2015}}

%
\begin{abstract}
We present a new model for reliability analysis that is able to
distinguish the \textit{latent} internal vulnerability state of the
equipment from the vulnerability caused by temporary \textit{external}
sources. Consider a wind farm where each turbine is running under the
external effects of temperature, wind speed and direction, etc. The
turbine might fail because of the external effects of a spike in
temperature. If it does not fail during the temperature spike, it could
still fail due to internal degradation, and the spike could cause (or
be an indication of) this degradation. The ability to identify the
underlying latent state can help better understand the effects of
external sources and thus lead to more robust decision-making. We
present an experimental study using SCADA sensor measurements from wind
turbines in Italy.
\end{abstract}

%
\end{frontmatter}

\section{Introduction}

One of the most important decisions that many companies face is when to
turn off mechanical equipment in order to perform preventive
maintenance. Considering wind farm maintenance, for instance,
it is much more cost effective to shut a turbine off before it fails
than to repair extensive damage caused by a failure. The goal then
becomes one of prediction: if we stop the turbine too early before it
would have failed, we lose valuable operating time. If we stop it too
late, the turbine may have sustained a catastrophic failure that is
expensive to repair. While the equipment is operating, its
vulnerability to failure depends not only on external factors such as
temperature, wind direction and speed, but also on latent degradation
due to wear-and-tear. If it can be well estimated, this latent
vulnerability state would be important to decision-makers because: (i)
it would provide insight into the health state of the equipment without
the influence of additional external factors, (ii) it would determine
whether the turbine is likely to sustain extreme external conditions
such as high temperatures, (iii) it would reveal how the various
external factors influence the degradation levels of the equipment, and
(iv) it would help provide maintenance decision-makers with a tool that
can help prevent too many early or late warnings. Ideally, we would
like to decouple the latent vulnerability state from the vulnerability
due to external sources, making as few distributional assumptions as possible.

Like many other types of large mechanical equipment (e.g., oil drilling
equipment, electrical feeders), wind turbines are usually equipped with
supervisory control and data acquisition (SCADA) sensors that record
various measurements of the dynamic environment every few minutes. For
each of the 28 turbines in Italy that we are considering, there are
over 30 different measurements collected every 10 minutes, including
temperatures inside and outside the turbine, wind speed, measurements
of the yaw system, and so on. These can be environmental measurements,
measurements that reflect the degradation state, or summary statistics
of these measurements. Wind farm operators make critical decisions on a
regular basis that depend on condition monitoring; this is because
failures occur fairly frequently, once every 2--6 weeks on average. The
frequency of these failures makes statistically motivated
decision-making strategies very relevant.

Our model is a generalization of the Cox proportional hazard model
(PHM) in that there are two separate terms in the hazard function, the
latent hazard $\mu$ and the transient hazard $g$. The main aspects of
the model can be summarized as follows:
\begin{longlist}[(4)]
\item[(1)] \textit{Latent State (Degradation) Term $\mu$}: This term is
monotonically nondecreasing over time, reflecting the fact that
mechanical equipment like a turbine does not self-heal. This term takes
into account the full history of the turbine as an integral of the
degradation. In contrast, to include the full history in the regular
Cox PHM, one would need to include a large number of terms and
constrain their influence to prevent the appearance of the nonphysical
self-healing.

\item[(2)] \textit{Transient Hazard Term $g$}: The second element of the
hazard function ($g$) reflects the instantaneous contribution to
vulnerability due to current measurements. For instance, a spike in
temperature would be reflected as a spike in the general $g$ term and
as a permanent rise in the $\mu$ term.

\item[(3)] \textit{Form of the Model}:
Our model generalizes the Cox proportional hazard model, which uses
only the $g$ term.
Our model is a mixed hazard model [see, e.g., \citet{Lin1995}]
whose form permits data-driven parameter fitting using convex
optimization techniques.

\item[(4)] \textit{Use in Decision-Making for Maintenance}:
Because we can isolate the latent state, we can better estimate the
resilience of our equipment: the latent state would not be as sensitive
to sudden but normal changes in the covariates. By triggering warnings
using knowledge of the latent state, decision-makers may be able to
issue better-timed warnings and alarms. We also provide an optimization
procedure to assist with decision-making.
\end{longlist}

The paper is organized as follows:
Section~\ref{secrelatedwork} includes the motivation for this work and
reviews the literature on using analytical techniques for
covariate-dependent degradation models, particularly for wind turbines
using SCADA data. Section~\ref{secmainsection} describes the model and
its derivation, along with an optimization procedure for making
maintenance decisions. Section~\ref{secproperties} discusses important
properties and inference. Section~\ref{secapplications} shows the
result of applying the proposed model and the warning generation
technique on turbines in Italy. Section~\ref{secnumericalexp} provides
a set of numerical experiments, including a simulation study and
comparisons with previous models. In the Supplementary Materials
\citet{Moghaddasssup}, we provide an interpretation of our model
and motivation with respect to discrete multi-state degradation models.

\section{Related work}\label{secrelatedwork} The literature of
degradation monitoring and failure analysis using time-varying
covariates can be divided into three main categories, namely, (1)
degradation-only modeling, (2) hazard modeling with diagnostic
covariates, and (3) degradation modeling with partial information. Our
work is related to all three categories.

Degradation-only models assume that the covariates are noisy versions
of the degradation state itself
[e.g.,
\citeauthor{Bian2012974} (\citeyear{Bian2012974,Bian2013}),
\citet{Gebraeel2008539},
\citet{Flory20141227},
\citet{Kharoufeh2003237},
\citet{Kharoufeh2005533},
\citet{Zhou20111586}].
In this way, one can assume, for instance, that each covariate could be
generated separately from the degradation state plus random noise. Some
of these studies have assumed that a failure occurs precisely when the
degradation signal exceeds a predefined threshold [e.g., \citeauthor{Bian2013} (\citeyear{Bian2012974,Bian2013}),
\citet{Gebraeel2008539}].
For example, \citet{Bian2013} presented a stochastic modeling
framework for sensor-based degradation signals for systems operating
under a time-varying environment. They assumed that the rate of
degradation directly depends on the environment profile that is known,
deterministic and evolves continuously. The overall degradation signal
is defined as the sum of the effect of environmental factors and a
time-dependent Brownian motion process.

For the case of wind turbines, it is not realistic to assume that we
simply have noisy measurements of the degradation state. Our
measurements all stem from (a possibly complicated) combination of
external sources of vulnerability and the degradation state itself
(e.g., temperature within the turbine); it is our job to separate these
two sources, and we cannot assume that the signals are directly
correlated with the underlying physical degradation processes.
In our case, the degradation process is assumed to be unobservable, and
there are no prior distributional assumptions on the parameters of the
model. We also do not require a predefined failure threshold, which is
often not available in real-world systems.

Works within the second group (hazard modeling with diagnostic
covariates) assume that the hazard rate is influenced by internal
and/or external time-varying covariates and aim to estimate the hazard
rate. The Cox proportional hazard model (PHM) [\citet{Cox1972}],
its time-dependent version [\citet{Fisher1999}] and its other extensions
[see, e.g., \citet{Gorjian2009}]
are examples of this second group. Works in the second group primarily
aim to predict the hazard rate, and do not necessarily model
degradation. Those that do model degradation generally take the
perspective of the first category, where some of the signals are known
to be noisy versions of the degradation state
[e.g., \citet
{Jardine1987,Qian2014317,Banjevic2006,Banjevic2001,Zhao2010,Wu2011,MAKIS1991}].

Some works consider Markov failure time processes
[e.g., \citet{Banjevic2006,Banjevic2001}]
where a predetermined subset of the covariates changes over time.
Finite state space models are useful in that the vulnerability states
are finite and meaningful, but the assumption of a finite state space
is restrictive and not particularly realistic. At the same time,
relaxing the assumptions of finite state space and considering multiple
covariates
yield too many states and transitions, making it difficult to estimate
all of the transition probabilities.

The model in this work, by contrast, does not require the number of
states to be known a priori, and the unobservable vulnerability state
is modeled as a function of past measurements. Furthermore, the model
in this work does not prespecify which variables contribute to the
degradation state, allowing this to be learned from data.

Our model is a particular time-dependent, additive-multiplicative mixed
hazard model (AMMHM), containing both additive and multiplicative
terms. There are examples of additive hazard models [e.g., \citet
{Pijnenburg1991}] and multiplicative hazard models
[e.g., \citet{Kalbfleisch2002}] in different application domains.
A review of hazard models with covariates, with an explanation of
AMMHM, is given by \citet{Gorjian2009}. Few studies have developed
special-purpose mixed hazard models. \citet{Martinussen2002}
proposed an additive-multiplicative model consisting of two components.
The first component contains additive covariates through an additive
Aalen model and the second component contains a multiplicative
covariate effect through a Cox regression model. Two different feature
sets were used to separately model baseline mortality and dose effects.
\citet{Anderson1989} considered a mixed model with additive and
multiplicative terms, where one term is proportional to a known
population mortality. They illustrated their model by predicting
survival of medical patients after an operation for malignant melanoma.
Another example of a mixed hazard model was proposed by \citet
{Hoehle2009} for spatial and temporal infectious disease surveillance.
For a theoretical analysis of AMMHM, interested readers may refer to
the work of \citet{Lin1995}. Our model leverages the
multiplicative and additive terms for a specific purpose: to separate
the latent hazard state from the external risk factors. Our particular
multiplicative term $(\mu)$ acts as a concise representation of the
full history of covariates. It is unclear how one would encode the full
history of covariates within, for instance, the Cox proportional
hazards model, without introducing a large number of variables.

The third group of related models are partially observable degradation
models. These models differ from the other two types in that the
degradation process is assumed to be unobservable (hidden). Instead,
covariates with an indirect relationship to the degradation process are
monitored over time. These are also referred to as degradation
processes with incomplete information [\citet{Hontelez1996267}] or
partially-observed degradation processes [\citet{Ghasemi2007989}].
Most works in this category have used hidden Markov and hidden
semi-Markov models with discrete unobservable degradation states
[\citet{Peng2011237,Moghaddass201294}]. Our model is similar to
those in this category in that it also provides insight into the latent
degradation state. Our model is different from those in this category
in that we do not need to specify the number of states, the transition
probability distribution between states or the structure of the
stochastic relationship between the covariates and the degradation
process. The covariates used in our work, by contrast, are not
necessarily assumed to reflect the hidden degradation level.

There are other types of degradation models that make particular
generative assumptions for specific types of environmental processes.
For example, \citet{Kharoufeh2006303} considered a single-unit
degrading system affected by its operating environment with a
deterministic degradation threshold value. They considered a random
shock process where the rate of wear is modulated by a discrete-space,
continuous-time Markov chain, and additional damage is induced by a
Poisson shock process. The total degradation is assumed to be the sum
of the degradation due to wear and that due to shocks. For other types
of degradation models used in reliability modeling, interested readers
may refer to the works of \citet{Gorjian2009} and \citet{Si20111}.

For wind farms, keeping maintenance costs low is essential; it is
difficult to be competitive against the costs of other energy sources,
such as oil and gas. There are several reviews on condition monitoring
and fault detection at wind farms [e.g., \citet
{Hameed2009,Kusiak2013,Lu2009,Marquez2012}]. In the wind industry,
SCADA (supervisory control and data acquisitions) systems are the most
commonly used mechanism for turbine health monitoring [\citet
{Marquez2012}]. Although SCADA systems are relatively inexpensive to
install and are used at almost all wind farms, relatively very little
research effort has been devoted to analytics using wind turbine SCADA
measurements. There are several potential reasons for this. For
instance, one is that SCADA sampling frequency is too low to be used
for spectral analysis, and another is that it does not collect all the
information needed for full condition monitoring of any particular wind
turbine component. On the other hand, since SCADA measurements do
provide ample and cheap indirect information about the health state of
the turbine, some research has begun to determine how to leverage these
measurements for health monitoring. \citet{Qiu2012} proposed an
alarm analysis and prioritization methodology using descriptive
statistics of SCADA data. A method for processing SCADA data and a
condition monitoring technique were developed by \citet{Yang2013}
using a regression approach to anomaly detection. Several papers
[e.g., \citet{Zaher2009,Marvuglia2012574}] proposed 
using machine learning techniques for condition monitoring using
SCADA data. \citet{Guo2009} proposed a time-dependent reliability
analysis based on the three-parameter Weibull distribution for wind
turbine failure time data.
A recent review of challenges for wind turbine maintenance was provided
by \citet{Yang2014}.

To summarize, the benefits of our model beyond those of previous works
are that (i) it decouples the (unobserved) degradation state from the
hazard due to transient sources, without having to specify anything
about the relationship of the features to the degradation state, (ii)
it takes the full history of measurements into account in a concise
way, which cannot be done easily in a Cox proportional hazard model
without including a large number of terms, (iii) it provides a new
decision-making methodology through latent state inference.

\section{The latent state hazard model}
\label{secmainsection}
Although this work focuses on turbine modeling, our approach can be
applied to any type of nonself-healing degrading system. Our notation
is as follows:
\begin{longlist}
\item[$N$:] Total number of units (turbines).

\item[$P$:] Number of features (SCADA measurements) used for monitoring.

\item[$T_i$:] The total lifetime of the $i$th unit.

\item[$\Delta$:] Measurement interval.

\item[$x_{i,k}(j)$:] The value of the $k$th covariate at time $j\Delta$ for
the $i$th unit.

\item[{$\mathbf{x}_i(j)=[x_{i,1}(j),x_{i,2}(j),\ldots,x_{i,P}(j)]^\top$:}] Feature
measurements at time $j\Delta$ for the $i$th unit.

\item[{$\mathbf{x}^h_i(j)=[\mathbf{x}_i(1),\mathbf{x}_i(2),\ldots,\mathbf{x}_i(j)]$:}] History
of features up to time $j\Delta$ for the $i$th unit.
\end{longlist}

We will assume, mainly for notational convenience, that hazard rates
are constant over each small unit of time $\Delta$ and can be presented
as piecewise constant functions of time. The hazard rate in each
interval can thus be approximated by the hazard rate at the endpoint of
that interval. Given the full history of covariate values $\mathbf
{x}_i^h(t)$, the notation for the hazard rate is $\lambda (t\mid\mathbf
{x}_i^h(t) )$, which we model with two terms as follows:
\begin{equation}
\label{eqhazardfunctionmain} \lambda\bigl(t\mid\mathbf{x}_i^h(t)\bigr)=\mu
\bigl(t\mid\mathbf{x}_i^h(t)\bigr)+g\bigl(t\mid
\mathbf{x}_i(t)\bigr),
\end{equation}
where, assuming $t$ is a multiple of $\Delta$, we set
\begin{eqnarray}
\label{eqinternalmain} \mu\bigl(t\mid\mathbf{x}_i^h(t) \bigr)&=&
\int_{0}^{t}\mu_0(\tau) \exp\bigl(
\beta_0+\bolds{\beta}^\top\mathbf{x}_i(\tau)
\bigr) \,d\tau\nonumber
\\
&\approx& \sum_{l \in\{\Delta,2\Delta,3\Delta,\ldots,t\}}\mu_0(l)
\exp\bigl(\beta_0+\bolds{\beta}^\top\mathbf{x}_i(l)
\bigr)\Delta
\\
\nonumber
&=&\mu\bigl(t-\Delta\mid\mathbf{x}_i^h(t-
\Delta) \bigr) + \mu_0(t) \exp\bigl(\beta_0+\bolds{
\beta}^\top\mathbf{x}_i(t) \bigr)\Delta.
\end{eqnarray}
%
The $\mu$ term is the latent state term. Because it is an integral of
exponentials, it is monotonically nondecreasing in $t$. The
approximation in the second line of equation \eqref{eqinternalmain}
shows the discrete form of the integral (which we use in practice since
our measurements are taken at discrete times), and the third line in
equation \eqref{eqinternalmain} shows that it can be written as a
recursion. The other term of $\lambda$ is the $g$ term, referred to in
this paper as the transient hazard term, which is not necessarily
monotonically increasing. The mathematical expression of the $g$ term is
\begin{equation}
\label{eqexternalmain} g \bigl(t\mid\mathbf{x}_i(t) \bigr)= g_0(t)
\exp\bigl(\alpha_0+\bolds{\alpha}^\top
\mathbf{x}_i(t) \bigr).
\end{equation}
In our notation, the feature vector $\mathbf{x}_i(t)$ depends on time $t$,
but $\mathbf{x}_i^h(t)$ could have components that are feature values from
the previous times, or nonlinear transformations of measurements taken
either at the current time or in the past. The coefficients $\beta_0$
and $\alpha_0$ are intercept terms and $ \bolds{\beta}=\{\beta_1, \ldots,\beta_P\}^\top$ and $\bolds{\alpha}=\{\alpha_1, \ldots,\alpha_P\}
^\top$ are each the vector of regression coefficients associated with
$P$ features in $\mu$ and $g$, respectively. The functions $\mu_0(t)$
and $g_0(t)$ are, respectively, the baseline hazard functions associated
with $\mu$ and $g$. We note here that $\beta_0$ and $\alpha_0$ could be
absorbed into the baseline hazard functions. Similarly, $\mu_0(t)$ and
$g_0(t)$ could be absorbed into the exponential term. The function $\mu
$ depends on the full history of covariates and encodes the latent
hazard state. We conjecture that $\mu$ is often smooth, but will
include jumps in the presence of fast-changing external factors to the
degradation state. The $g$ term should similarly fluctuate as a
function of current external conditions (e.g., temperatures and wind
speeds). It is possible to make the transient hazard term $g$ depend on
$\mu$ by adding terms within $g$ related to $\mu$.
The main parameters of the model are the intercepts $\alpha_0$ and
$\beta_0$, and the $\bolds{\alpha}$ and $\bolds{\beta}$ vectors, each of size
$P$, which encode the history of the degradation process and the
temporary influences of the covariates.

\begin{figure}

\includegraphics{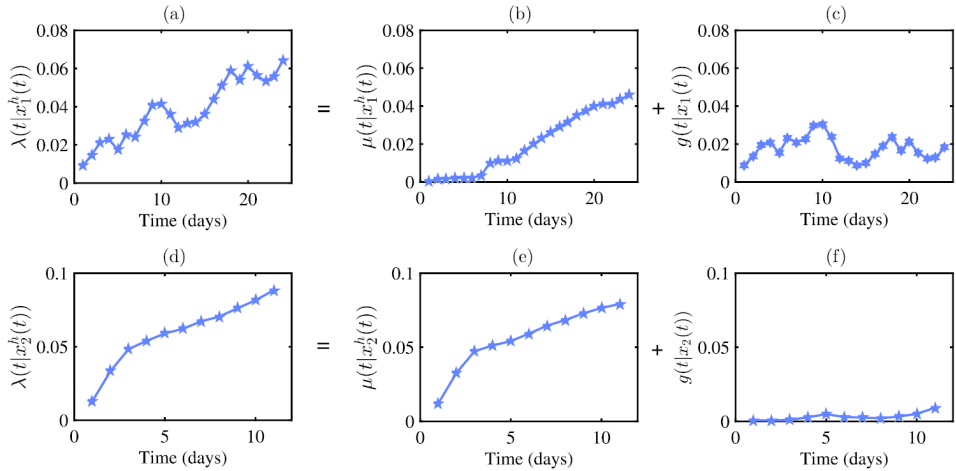}

\caption{Estimated hazard rates ($\lambda$, $\mu$ and $g$ in columns
1--3, resp.) for turbine No. 1 (1st row) and turbine No. 2 (2nd
row), decomposed into the latent degradation ($\mu$) and the transient
vulnerability ($g$), using the model presented in this study.
\textup{(a)}~Turbine No.~1 $(\lambda)$,
\textup{(b)}~turbine No.~1 $(\mu)$,
\textup{(c)}~turbine No.~1 $(g)$,
\textup{(d)}~turbine No.~2 $(\lambda)$,
\textup{(e)}~turbine No.~2 $(\mu)$,
\textup{(f)}~turbine No.~2 $(g)$.}
\label{figlshm}
\end{figure}

In expressions \eqref{eqinternalmain} and \eqref{eqexternalmain} the
covariates, such as temperature, pressure, etc., are not assumed to
follow a certain distribution or well-structured time series.
Figure~\ref{figlshm} shows an example of the total hazard rate, latent
vulnerability rate $\mu$ and transient vulnerability $g$, that we
estimated from two of the wind turbines. For the first turbine shown in
Figure~\ref{figlshm}(a)--(c), the total hazard rate is formed from a
fairly balanced mix of both degradation and temporary sources. For the
second turbine shown in Figure~\ref{figlshm}(d)--(f), the total hazard
rate comes mainly from degradation, and there is minimal contribution
from transient sources. The model parameters used for both turbines are
the same, and were learned from a separate training set that did not
include either of the turbines.
%

\subsection{Cost-benefit analysis and decision-making}\label
{seccostbenefit}
\begin{figure}[b]

\includegraphics{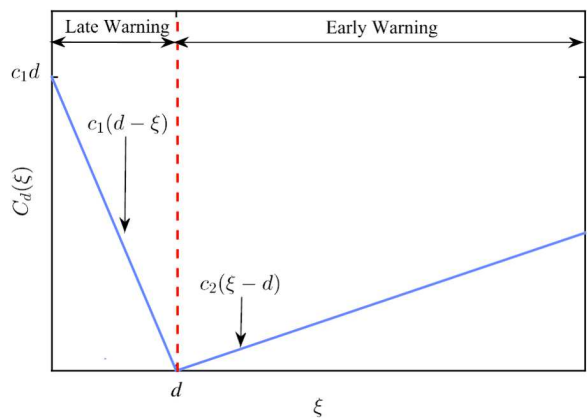}

\caption{An example of the function $C_d(\xi)$, as explained in the
text. The cost of warning at exactly $d$ units before failure is zero,
and there is a positive cost for warnings that are too early (i.e., for
$\xi>d$) or too late (i.e., for $\xi<d$). The unit costs of late
warning and early warning are $c_1$ and $c_2$, respectively.}
\label{figsamplecdq}
\end{figure}

We propose to use our model for generation of warnings.
It is important that our warning generation method is accurate. For
instance, alarm and warning rates that are too conservative can not
only increase downtime and decrease productivity, but they can also
reduce the operators' sensitivity to failure, which can have
catastrophic consequences. We can define ``warning generation'' as a
decision process that depends on the estimated hazard of each unit over
time. Let us define $d$ as the ideal lead time between the warning
point and the failure point determined by decision-makers. In other
words, the warning generation system is considered efficient if it
generates warnings when the actual time to failure (also called
remaining useful life---RUL) is very close to $d$ time units. To find
the optimal warning policy, we can define $\gamma_{d}$ as the threshold
for warning generation, so that a warning is put in place as soon as
the estimated hazard exceeds this threshold. To define the quality of
the decision framework, we define a cost function $C_d(\xi), \xi\geq
0$, to represent the cost of a warning at $\xi$ units before the actual
failure time. We should note that $C_d(d)=0$, and there is a positive
cost for warnings that are too early or too late. The cost of an actual
failure without warning in advance, that is, when the warning time
equals the failure time, is $C_d(0)$. We refer to this cost as the cost
of warning at failure. Our model is general in the sense that any kind
of cost function can be considered depending on the application (e.g.,
hinge, quadratic, logistic, exponential, etc.). A very simple form of
this function is shown in Figure~\ref{figsamplecdq} and is typically
called the ``pinball loss'' or ``newsvendor cost'' [see, e.g.,
\citet{Rudin2014}].

%
We define $R_{i,\gamma_d}$ as the model's warning time for the $i$th
available lifetime based on the lead time $d$ and threshold $\gamma
_{d}$. For any $d$, one can determine the optimal policy $\gamma^*_{d}$
such that the expected average cost of the warning generation process
is minimized. We find the optimal solution $\gamma^*_{d}$ using an
empirical risk model (ERM). The optimization problem is
\begin{eqnarray}
\label{eqwarninggenerationprocess}
\mathop{\operatorname{argmin}}_{\gamma_d}
J_d(\gamma_d)\qquad\mbox{where } J_d(
\gamma_d) = \frac{1}{\llvert N_1\rrvert }\sum_{i \in N_1, T_i \geq d}
{C_d(T_i-R_{i,\gamma_d})},
\nonumber\\[-8pt]\\[-8pt]
\eqntext{\displaystyle\mbox{where }
R_{i,\gamma_d} = \min
\bigl\{ T_i, \inf\bigl\{ j\mid\hat{\lambda}\bigl(j\mid
\mathbf{x}^h_i(j)\bigr)\geq\gamma_d, j \geq0
\bigr\} \bigr\}, i \in N_1,  \gamma_d \geq0,}
\end{eqnarray}
where $N_1$ are the lifetimes used for training,
$\inf \{ j\mid \hat{\lambda} (j\mid\mathbf{x}^h_i(j) )\geq\gamma_d, j
\geq0 \}$ is the time at which the estimated hazard rate ($\hat{\lambda
}$) of the $i$th unit exceeds the threshold, and $\gamma_d$ is the only
decision variable. The warning time $R_{i,\gamma_d}$ is either the time
at which the unit fails ($T_i$) or the time at which its estimated
hazard rate exceeds the threshold $\gamma_d$, whichever occurs first.

\begin{figure}

\includegraphics{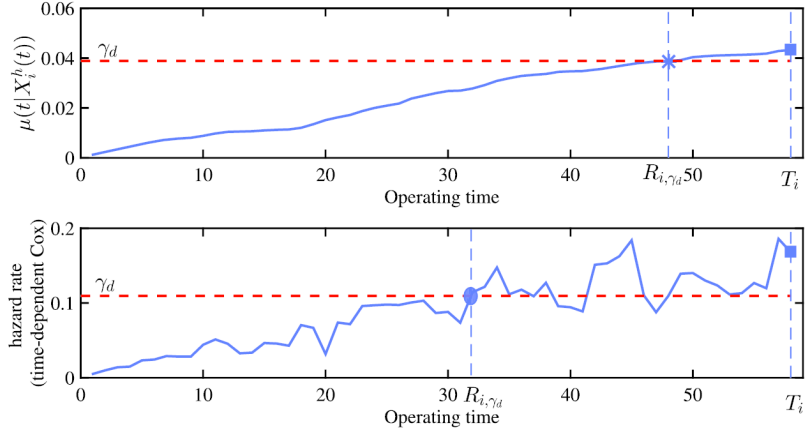}

\label{figsamplecd}
\caption{Hazard rates (solid lines), decision thresholds ($\gamma_d$,
dashed lines), warning times ($R_{i,\gamma_d}$) and failure times
($T_i$) for a single lifecycle of a turbine. The hazard rates have been
estimated using (top) the $\mu$ term from the latent state hazard model
and (bottom) the hazard rate from the Cox proportional hazard model.
The decisions based on the latent state tend to be more robust to
fluctuations earlier on in the timeline, possibly leading to more
accurate lifecycle estimates.
}
\label{figpolicies}
\end{figure}

Figure~\ref{figpolicies} shows how decisions can be made using the
latent state $\mu$ for our model, in contrast with how decisions would
be made using the (more common) time-dependent proportional hazard
model, where hazard rate is employed as the decision criteria. This
figure also illustrates why decisions made using the latent degradation
state $\mu$ tend to be more robust; they are resilient to fluctuations
in the hazard rate earlier in the lifecycle of the turbine. This can
lead to longer (more accurate) lifecycles and thus lower costs.
%

\section{Properties and inference}
\label{secproperties}
In this section we discuss how to fit the model to data.
\subsection{Model training}\label{subsecmodeltraining}
We use the method of maximum likelihood to infer the coefficients of
the model. Given lifetimes with stopping times $T_i$ for each turbine~$i$, the continuous time version of the likelihood is
\begin{equation}
\mathcal{L}_N(\bolds{\theta}\mid X)=\prod_{i=1}^N
{ \biggl(\exp\biggl(-\int_{0}^{T_i}{\lambda\bigl(t
\mid\mathbf{x}_i^h(t)\bigr)} \,dt \biggr) \biggr)} \times{
\lambda\bigl(T_i\mid\mathbf{x}_i^h(T_i)
\bigr)},
\end{equation}
where $\lambda$ depends on $\bolds{\theta}=\{\alpha_0,\bolds{\alpha},\beta
_0,\bolds{\beta}\}$ through the definition of the model in equations (\ref
{eqhazardfunctionmain}), (\ref{eqinternalmain}) and (\ref{eqexternalmain}).
If the lifetime associated with unit $r$ ($r \in\{1,2,\ldots,N\}$) is
censored at time $S_r$ $(S_r < T_r)$, then its contribution to the
likelihood function is
$\exp (-\int_{0}^{S_r}{\lambda(t\mid\mathbf{x}_r^h(t))} \,dt )$. From now
on, we assume all lifetimes are complete. For discretized time units,
the likelihood has the following form:
\begin{equation}
\label{eqlikelihoodmain} \quad\mathcal{L}_N(\bolds{\theta}\mid X)=\prod
_{i=1}^N \underbrace{ \prod
_{j=1}^{T_i-1} \exp\bigl(- \lambda\bigl(j\mid
\mathbf{x}_i^h(j) \bigr)\Delta\bigr) }_{\mathrm{probability~of~survival}}
\times
\underbrace{ 
\bigl(1-\exp\bigl(- {\lambda\bigl(T_i\mid
\mathbf{x}_i^h(T_i) \bigr)} \Delta\bigr) \bigr)
} _{\mathrm{probability~of~failure}},
\end{equation}
where $\Delta$ is the length of each measurement interval. For
mathematical convenience, we let $\Delta=1$ and also define an
indicator variable $y_{i,j}$ as
\begin{equation}
y_{i,j}= \cases{ 1, &\quad${j}=T_i$,
\cr
0, &\quad$ {j}
\neq T_i$.}
\end{equation}
Then the log-likelihood function may be written as follows:
\begin{eqnarray}
\log\mathcal{L}_N(\bolds{\theta}\mid X) & =&\sum
_{i=1}^{N} \sum_{j=1}^{T_i-1}
-\lambda\bigl( j\mid\mathbf{x}_i^h (j ) \bigr) + \sum
_{i=1}^{N} \log\bigl( 1 - \exp\bigl( -
\lambda\bigl(T_i\mid\mathbf{x}_i^h
(T_i ) \bigr) \bigr) \bigr)
\nonumber\\[-8pt]\\[-8pt]\nonumber
&=&\sum_{i=1}^{N} \sum
_{j=1}^{T_i} \log\bigl( y_{i,j}+(1-2
y_{i,j})\exp\bigl( - \lambda\bigl(j\mid\mathbf{x}_i^h
({j}) \bigr) \bigr) \bigr),
\end{eqnarray}
where $\bolds{\theta}$ is the set of parameters of the model. It is now
clear that we have
\[
\hat{\bolds{\theta}}\in\arg\mathop{\max}_{\bolds{\theta}} \log\mathcal{L}_N(
\bolds{\theta}\mid X),
\]
where $\hat{\bolds{\theta}}$ is the maximum likelihood estimate of the
parameters. We will show later that, under some regularity conditions,
$\hat{\bolds{\theta}}$ converges to the true parameter set $\bolds{\theta}^0$
in probability when $N$ is large. The procedure outlined above will
produce a point estimate for $\bolds{\theta}$. (If full Bayesian inference
is desired, we could assume a normal prior on $\bolds{\theta}$ and sample
from the posterior distribution over $\bolds{\theta}$. However, this would
lose the interpretability of the single point estimate, be far less
computationally tractable, and the mechanism for making decisions using
the full posterior would likely require us to choose a point estimate anyway.)

\subsection{Regularization}
We use $\ell_2$ regularization or, equivalently, a normal prior on
model parameters $\bolds{\alpha}$ and $\bolds{\beta}$. Regularization helps
prevent overfitting, and makes the log of the objective function
strictly convex. In particular, we optimize
\begin{eqnarray}\label{eqlossfunction}
W_N(\bolds{\theta}\mid X) &=& - \log\mathcal{L}_N(
\bolds{\theta}\mid X)+ \mathbf{C}\llVert\bolds{\theta}\rrVert_2^2
\nonumber\\[-8pt]\\[-8pt]\nonumber
&=& - \log\mathcal{L}_N(\bolds{\theta}\mid X)+ C_1 \llVert
\bolds{\alpha}\rrVert^2_2 +C_2\llVert\bolds{\beta}
\rrVert^2_2.
\end{eqnarray}
Setting $C_1$ very large will cause the model to ignore the internal
state. Similarly, setting $C_2$ very large will result in ignoring the
transient hazard term $g$. In practice, we set $C_1$ and $C_2$ using
cross-validation; however, they can be set manually, to force more
weight to internal degradation or vice versa. As usual, the
regularization constants should effectively vanish as $N$ tends to infinity.

\subsection{Convexity of the loss function}\label{secconvexity}

%
\begin{proposition}\label{11111}
The loss function $-\log\mathcal{L}_N(\{\bolds{\alpha},\bolds{\beta}\}\mid
X)+ C_1 \llVert \bolds{\alpha}\rrVert ^2_2 +C_2\llVert \bolds{\beta
}\rrVert ^2_2$ derived from equation (\ref{eqlossfunction}) is
strictly convex when $C_1>0$, $C_2>0$.
\end{proposition}

The proof is given in Appendix~\ref{app1}.
\subsection{Coordinate descent method for model training}\label
{secCoordinateDecent}
Since the optimization problem is convex and differentiable, coordinate
descent is a natural fit. The direction is provided by a Fr\'echet
(directional) derivative. Denoting $C_1 \llVert \bolds{\alpha}\rrVert
^2_2 + C_2 \llVert \bolds{\beta}\rrVert ^2_2$ by $\mathbf{C}\llVert \bolds
{\theta}\rrVert _2^2$, we have
%
\begin{eqnarray}
\label{eqgradient}
\qquad && \frac{\partial}{\partial\theta_k} \bigl[-\log
\mathcal{L}_N(\bolds{\theta}\mid X)+ \mathbf{C} \llVert\bolds{\theta}\rrVert^2_2\bigr] \nonumber
\\
&& \qquad =
\sum_{i=1}^{N} \sum
_{j=1}^{T_i} (1-2y_{i,j}) \frac{\exp(-\lambda({j}\mid\mathbf{x}_i^h({j})) )}{
y_{i,j}+(1-2y_{i,j})\exp(-\lambda({j}\mid\mathbf{x}^h({j})) )}
\frac{\partial\lambda({j}\mid\mathbf{x}_i^h({j}))}{\partial\theta_k}
\\
&&\quad\qquad{} + \frac
{\partial}{\partial\theta_k} \mathbf{C} \llVert\bolds{\theta}\rrVert
^2_2,\nonumber
\end{eqnarray}
where $\theta_k$ is the $k$th parameter of $\bolds{\theta}$, and
\begin{equation}
\frac{\partial}{\partial\theta_k} \lambda\bigl({j}\mid\mathbf{x}_i^h({j})
\bigr)= \frac{\partial}{\partial\theta_k} \mu\bigl({j}\mid\mathbf{x}_i^h({j})
\bigr)+ \frac{\partial}{\partial\theta_k}g \bigl({j}\mid\mathbf{x}_i({j})
\bigr).
\end{equation}
Now, for the coefficients of ${\bolds{\beta}}$ (denoting $\theta_k = {\beta
}_{k_1}$), we have
\begin{equation}
\frac{\partial}{\partial\beta_{k_1}}\lambda\bigl({j}\mid\mathbf{x}_i^h({j})
\bigr)=\sum_{l=1}^{j} x_{i,k_1}({l})
\exp\bigl(\beta_0+\bolds{\beta}^\top\mathbf{x}_i({l})
\bigr).
\end{equation}
Similarly, for the coefficients of $\bolds{\alpha}$ (denoting $\theta_k =
{\alpha}_{k_2}$), we have
\begin{equation}
\frac{\partial}{\partial\alpha_{k_2}}\lambda\bigl({j}\mid\mathbf{x}_i^h({j})
\bigr)=x_{i,k_2}({j})\exp\bigl(\alpha_0+\bolds{
\alpha}^\top\mathbf{x}_i({j}) \bigr).
\end{equation}
In Algorithm \ref{algalg1}, the steps for optimizing the loss function
using the Coordinate Descent method are described.

\begin{algorithm}[t]
\caption{Coordinate Descent Algorithm}
\label{algalg1}
Let $W_N(\bolds{\theta}\mid X) = - \log\mathcal{L}_N(\bolds{\theta}\mid X)+
\mathbf{C}\llVert \bolds{\theta}\rrVert _2^2$ as in equation \eqref
{eqlossfunction}, then:
%
\begin{enumerate}[2.]
\item[1.] Select the starting point $\bolds{\theta}^{(1)}$ and the convergence
parameter
$\varepsilon$ and let $k=1$.
\item[2.] Compute $\frac{\partial}{\partial\theta_i} [ W_N(\bolds{\theta
}^{(k)}\mid X) ] $ for all $i$ from equation (\ref{eqgradient}).
\item[3.] Choose $i_k\in\arg\mathop{\max}_{i} \mid\frac{\partial}{\partial
\theta_i} [ W_N(\bolds{\theta}^{(k)}\mid X) ]\mid$.
\item[4.] Find the positive step size ($\varphi_k$) as
%
\begin{equation}
\varphi_k \in\arg\min_{x} W_N
\biggl(\bolds{\theta}^{(k)} - x \frac{\partial}{\partial\theta_{i_k}} \bigl
[ W_N
\bigl(\bolds{\theta}^{(k)}\mid X\bigr) \bigr] e_{i_k}\mid X \biggr).
\end{equation}
Here, $e_i$ is the $i$th coordinate vector in $\mathbb{R}^{2P+2}$.
\item[5.] Update the current point as
\[
\bolds{\theta}^{(k+1)}=\bolds{\theta}^{(k)}- \varphi_k
\frac{\partial}{\partial\theta_{i_k}} \bigl[ W_N\bigl(\bolds{\theta
}^{(k)}\mid X
\bigr) \bigr] e_{i_k}.
\]
\item[6.] Evaluate $W_N(\bolds{\theta}^{(k+1)}\mid X)$. If the condition
$\mid W_N(\bolds{\theta}^{(k+1)}\mid X)-W_N(\bolds{\theta}^{(k)}\mid X)\mid
<\varepsilon$ is satisfied, then terminate the algorithm and output $\bolds
{\theta}^*=\bolds{\theta}^{(k+1)}$. Otherwise, set $k=k+1$ and return to
step~2.
\end{enumerate}
\end{algorithm}

\subsection{Asymptotic properties}
In this section we state a result (whose proof is in Appendix~\ref{app2}) about the consistency properties of the maximum likelihood
estimators of the parameters of our model.

\begin{theorem}\label{theoremconsis}
Let $X=(X_i, T_i), 1 \leq i \leq N$, be i.i.d. with the likelihood
function $\mathcal{L}_N(\bolds{\theta}\mid X)$ given in equation (\ref
{eqlikelihoodmain}) with independent parameter set $\bolds{\alpha}$ and
$\bolds{\beta}$ where $(\bolds{\alpha},\bolds{\beta})= \bolds{\theta} \in\Theta$,
$\mid\bolds{\alpha}\mid<M$, and $\mid\bolds{\beta}\mid<M$, where $M$ is a
finite positive constant independent of $N$. Then with probability
tending to 1 as $N$ tends to infinity, there exist solutions
$\hat{{\bolds{\theta}}}_N=
(\hat{{{\theta}}}_{N,1},\ldots,\hat{{{\theta}}}_{N,p})$ of the
likelihood equations such that:
\begin{longlist}[(iii)]
\item[(i)] $\hat{{{\theta}}}_{N,j}$ is consistent for estimating $\theta_j$,

\item[(ii)] $\sqrt{N} (\hat{{\bolds{\theta}}}_{N}-\bolds{\theta}^0)$ is
asymptotically normal with (vector) mean zero and covariance
matrix $[\mathbf{I}(\bolds{\theta})]^{-1}$, and

\item[(iii)] $\hat{{{\theta}}}_{N,j}$ is asymptotically efficient in
the sense that its variance attains the Cram\'er--Rao lower bound as $N$
goes to infinity.
\end{longlist}
\end{theorem}

The proof is a consequence of Theorems \ref{theoremcconditions} and \ref
{theoregularity}, which are given in Appendix~\ref{app2}.
\section{Application to wind turbine data}\label{secapplications}
Our collaboration is with Accenture (a consulting company) and ENEL
(Italy's largest power company).
Our data are from a wind farm in Europe that collects SCADA data from
$N=28$ (with $N$ defined as in Section~\ref{secmainsection})
pitch-regulated 2 mega-watt wind turbines. The original measurements
used in this work were collected every 10 minutes over the course of a
year, and values were averaged over the course of a day. Averaging
smooths out small variations and is much more computationally
efficient. However, too much averaging can potentially remove the
effect of $g$. In order to make sure that we prevent this from
happening, we repeated some of our experiments on smaller portions of
our data, averaging over 12-hour intervals and 6-hour intervals. The
estimates for $\mu$ and $g$ for these experiments were similar to those
for the 24-hour interval, implying that our choice of one day was reasonable.

\begin{figure}

\includegraphics{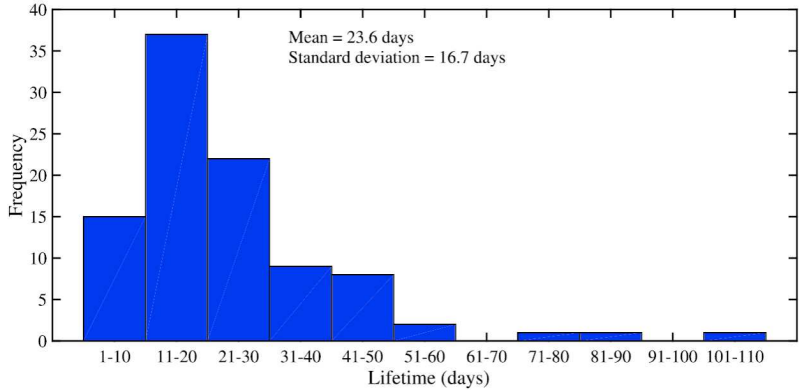}

\caption{Histogram for the 95 lifetimes associated with the 28
turbines.}\label{figlifetime}
\end{figure}

The original data consist of $P=49$ (with $P$ defined as in Section~\ref{secmainsection}) different signals including internal and external
covariates. External covariates are those generated by an independent
process that can influence (accelerate or decelerate) the degradation
state, such as environmental covariates (humidity, wind speed, external
temperature, time, etc.) and operational covariates (load). Also,
external covariates include the nontime-varying covariates, such as
turbine age in years, location and manufacturer. Internal covariates
are those relevant to estimating the degradation process. Examples are
oil temperature, gear box temperature and voltage. There were many
missing observations in the data, resulting from the SCADA sensors
being turned off or other SCADA malfunctions. We chose 95 lifetimes
that had few missing points to minimize the bias of the SCADA
malfunctions, though since we averaged data over each day, even a few
hours of missing points would not change our results very much. A
``lifetime'' is a time interval between the time point at which the
turbine is restarted after a work order ticket and when it fails. Thus,
there are possibly multiple lifetimes for each turbine. This implicitly
models the effect of maintenance as returning the turbine's condition
to a similar ``restored'' condition at the beginning of each lifetime,
though in reality we cannot always know the state of the turbine
exactly after a particular type of maintenance was performed.
We have assumed that these lifetimes are independent; if one has more
information about the state of the system after a repair, our model can
be modified to consider an initial degradation level [see \citet
{Moghaddasssup}]. Figure~\ref{figlifetime} displays the histogram for
95 lifetimes associated with the 28 turbines (also shown are the mean
and the standard deviation of the 95 lifetimes).

In addition to the modeling efforts discussed in this work, a lot of
the effort for this project went into the derivation of appropriate
features and labels. In particular, our SCADA data do not always
include an automated failure ``flag'' that indicates what part of the
turbine failed or whether the turbine had been shut off for reasons
other than failure (e.g., inspection). We used a separate database of
``work orders'' written by the wind farm company to help us determine
whether the turbine had been shut off due to failure; however, those
data were not sufficient to differentiate reasons for the shut-offs or
identify the particular parts of the turbines that failed. As a result,
we chose to predict whether any failure mode will occur, and thus
categorized the work orders into those that represented failures and
those that represented nonfailures. Table~\ref{tablefeaturelist}
provides the list of covariates we used and which part of the turbine
was measured to obtain each covariate. Since these covariates are
completely typical of data collected by SCADA systems for wind
turbines, we believe our approach could be widely applied by the wind industry.

%

%
\begin{table}[b]
\tabcolsep=0pt
\caption{Covariates selected for turbine health monitoring and the part
of the turbine used to measure them}\label{tablefeaturelist}
\begin{tabular*}{\tablewidth}{@{\extracolsep{\fill}}@{}lclc@{}}
\hline
\textbf{Feature name} & \textbf{Location} & \textbf{Feature name} & \textbf{Location}\\
\hline
Pitch average & Rotor & Bus voltage & Generator \\
Hydraulic pressure & Hydraulic & Temperature transformer max & Generator \\
Nacelle temperature & Nacelle & Temperature radiator 2 & Generator \\
Ambient temperature & Ambient & Temperature drive end bearing & Gearbox\\
Active power & Counter & Temperature nondrive end bearing & Gearbox \\
Power loss & Counter & Temperature gear oil & Gearbox \\
Temperature generator windings avg & Generator & Generator speed & Generator \\
\hline
\end{tabular*}
\end{table}

\begin{figure}

\includegraphics{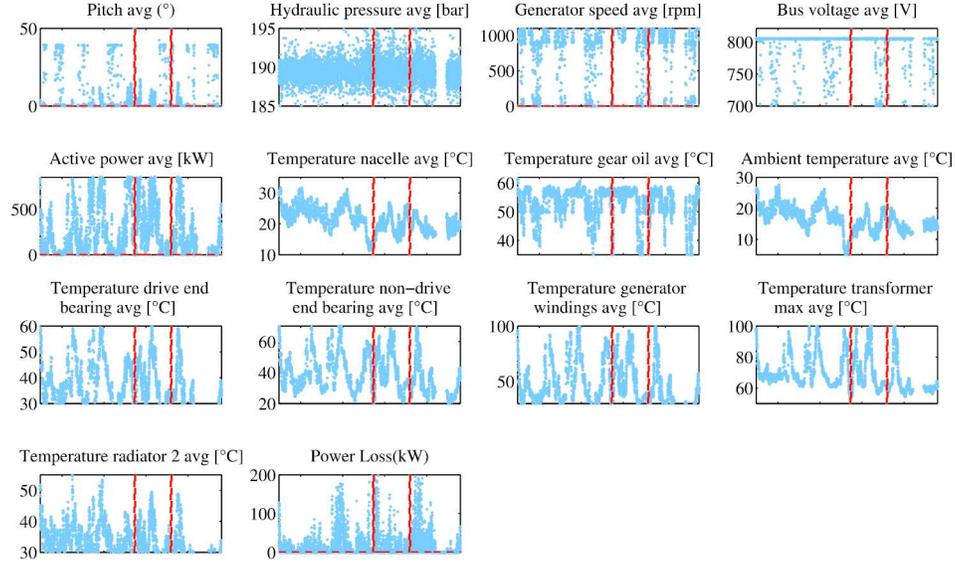}

\caption{Original feature values displayed for a period of 3 months for
a sample turbine. The dotted vertical lines in each plot represent work
order events (maintenance actions).}\label{figsamplefeaturevalues}
\end{figure}

In Figure~\ref{figsamplefeaturevalues}, the SCADA measurements of 14
covariates for a single turbine are plotted for a period of three
months. The dotted lines in each plot represent work order events
(maintenance actions). No obvious trend can be discerned from any
individual signal.
We normalized each signal to be between zero and one, where the minimum
and maximum used in the normalization were calculated over the training
set and the same values were used for the test set. This helps with
numerical stability and improves convergence speed for parameter estimation.

%
%

In addition to the features above, we added another class of features
that compares the signal values of each turbine to the other turbines
in the wind farm. For example, if the power output of one turbine is
much lower than the average generated power of the other turbines, it
could be an indication of a mechanical problem and a precursor to
failure. One class of features (denoted by $M_1$) is the differences in
percentile value of each signal from the median among other turbines
within the wind farm. That is, we subtracted 0.5 from the normalized
rank and took the absolute value to compute differences from the
median. The second class of features (denoted by $M_2$) is the
$z$-scores of the signal values. The formulas for these features are
given below:
\begin{eqnarray*}
M_1(i,j,t) &=& \biggl\llvert\frac{\sum_{j'} \mathbf{1}_{\{
x_{i,j}(t)<x_{i,j'}(t) \}}}{n(t)}-0.5 \biggr\rrvert,
\\
M_2(i,j,t) &=& \biggl\llvert\frac{x_{i,j}(t)-\bar{x}_{:,j}(t)
}{\operatorname{sd}(x_{:,j})(t)} \biggr\rrvert,\qquad
\forall(i,j,t),
\end{eqnarray*}
where $n(t)$ is the total number of turbines with nonnull measurements
at time $t$, $x_{i,j}(t)$ is the $j$th signal value associated with the
$i$th lifetime at time $t$, and $\bar{x}_{:,j}(t)$ and $\operatorname{sd}(x_{:,j})(t)$
are the average and the standard deviation of signal $j$ collected from
all available turbines at time $t$, respectively.

We used cross-validation to evaluate performance. We randomly divided
our data set into five folds of equal size (19 lifetimes per fold).
These sets are referred to as Test Set 1 (TS1) through Test Set 5 (TS5)
hereafter. We trained the model with 4 folds and used the last fold for
testing. This process was repeated 5 times so that all folds were used
for testing. We used Cox--Snell residuals [see, for more details,
\citet{Collett2003}] to check if the estimated hazard functions
model the set of turbines' lifetimes adequately. If the model fits the
data well, the Cox--Snell residuals should approximately follow a unit
exponential distribution. We used the Kolmogorov--Smirnov (K--S) test on
each test fold to compare the estimated cumulative hazard function
(Cox--Snell residuals) with the exponential distribution with mean 1.
Since all calculated $p$-values of the associated K--S test for the 5
folds are large (i.e., 0.14, 0.26, 0.30, 0.31, 0.19), we do not reject
the hypothesis that the model fits the data well. We should point out
that the power of the test increases rapidly away from a unit
exponential distribution with 19 lifetimes, which indicates that the
K--S test results are reasonable.

%
\begin{figure}

\includegraphics{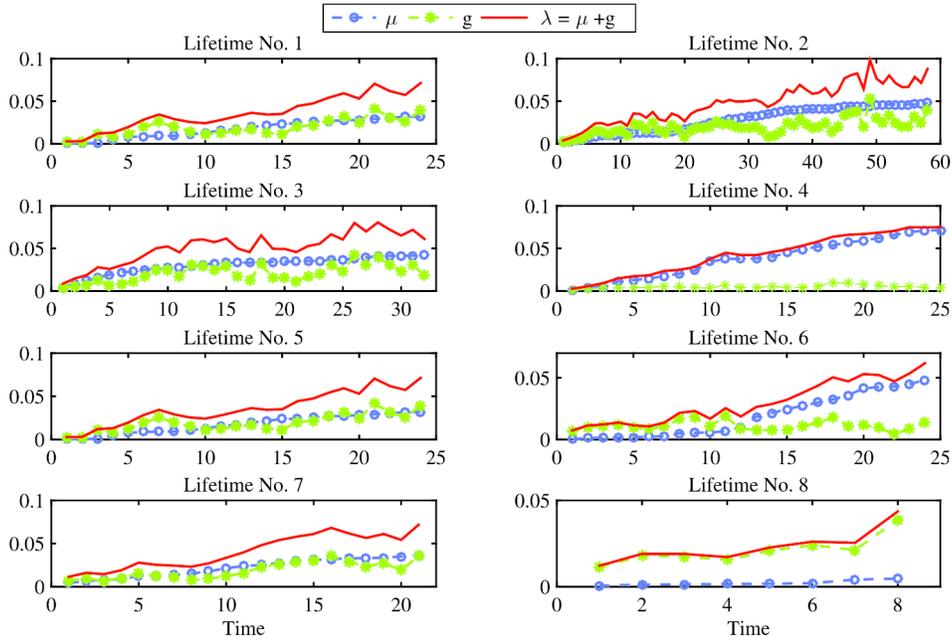}

\caption{Estimated $\lambda$ (solid lines), $\mu$ (dashed-star lines)
and $g$ (dashed-circle lines) for eight lifetimes in Test Set No.
1.}\label{figsamplefeaturevalues1}
\end{figure}

\subsection{Interpretability}
In Figure~\ref{figsamplefeaturevalues1}, the estimated hazard function
and its separation into $\mu$ and $g$ are shown for eight lifetimes in
the first test set. The model was trained on 76 lifetimes, using 0.1
for both regularization constants; the regularization constant was
found via a validation set of 19 training lifetimes.
The interesting feature of Figure~\ref{figsamplefeaturevalues1} is that
it shows a clear separation between internal and external effects,
where some of the lifetimes are driven mainly by the $\mu$ term, others
by the $g$ term, and some by both terms. Some of these effects may be
attributed to regularization, but not all. This means that some of the
lifetimes may have been more robust than others to external factors.
The increased robustness could be because the equipment is in a better
mechanical state and/or it is possible that the external effects are
mitigated due to the physical location of the turbine during that
lifetime. These possibilities may be explored further by the wind
turbine company, who may better be able to understand the cause of the
failures of the turbines and to use this information for planning
(locations of future turbines, maintenance policies, replacement
policies, etc.).

\subsection{Examining the hazard rate at the point of failure}
We wanted to know whether the hazard rates were high at the times when
the turbines actually failed.
We used rank statistics to do this.
In particular, for each lifetime, we evaluated the hazard rank
percentile when it failed. That is, we took each lifetime of length
($T_i$) and considered it a part of a cohort of all other lifetimes
equal to or exceeding $T_i$. Then at time $T_i$, we calculated the rank
percentile of the turbine that failed, which is the fraction of
turbines whose hazard rate was lower. The higher the percentiles, the
better our prediction method performed in terms of distinguishing
failures from nonfailures.

%
\begin{figure}

\includegraphics{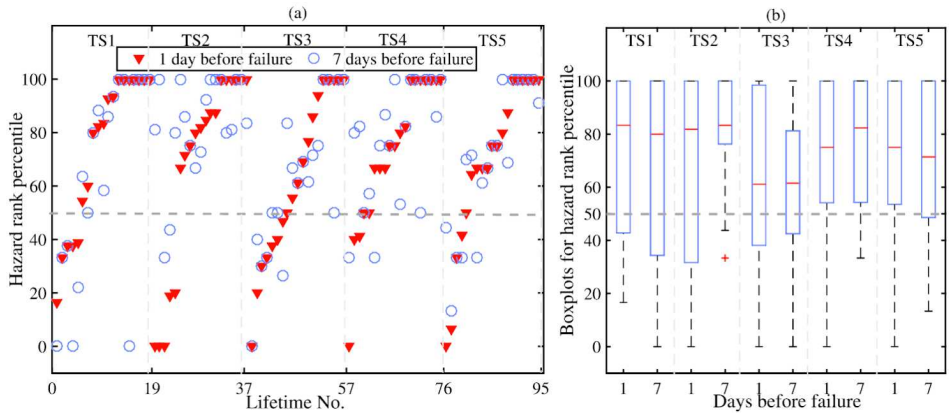}

\caption{\textup{(a)} Hazard rank percentile for the five test sets, with each
point representing the percentile rank of one turbine relative to all
working turbines with the values sorted by the percentile rank one day
before failure and \textup{(b)} the box plots for the rank percentile for the
five test sets; the first box plot in each test set is for 1 day before
failure. The dashed horizontal lines are located at the 50th
percentiles. In all test sets, most of the rank percentile distribution
is above the 50th percentile.}\label{fighazardrankpercentile}
\end{figure}

Figure~\ref{fighazardrankpercentile}(a) presents the hazard rank
percentile for all turbines in the five test sets (TS1--TS5) for one day
and one week before failure. Figure~\ref{fighazardrankpercentile}(b)
shows the same information, but in the form of box plots. It can be
observed from these figures that the hazard percentile of the failed
turbine is generally higher than those of the other operating turbines
with the same age. This is particularly true one day before the turbine
fails, but even one week before failure, the hazard rank percentile of
many turbines is still high, with the median percentile rank well above
50. This indicates that our model is performing well.

\subsection{Comparison between our model and the Cox proportional
hazard model}\label{secComparisonbetweenOurModelandCox}
\begin{figure}

\includegraphics{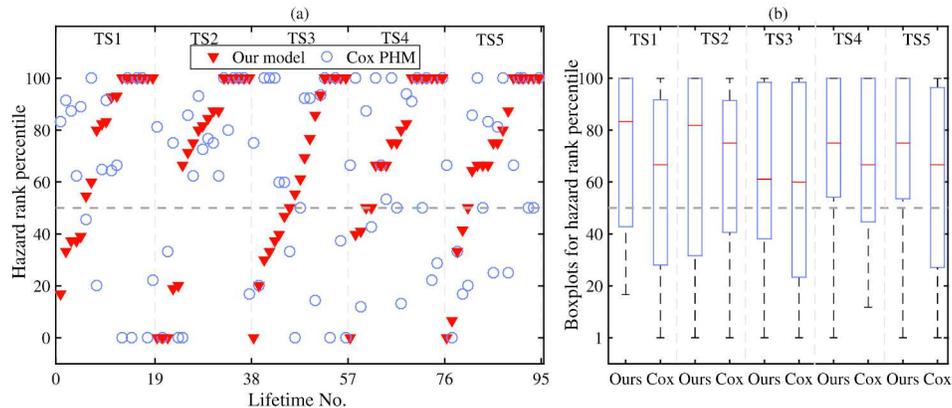}

\caption{Comparison between our model and Cox on the wind turbine data
set, \textup{(a)} hazard rank percentile for the five test sets, with each point
representing the percentile rank of one turbine relative to all working
turbines with the values sorted by our model and \textup{(b)} the box plots for
the rank percentile for the five test sets; the first box plot in each
test set is for our model. The dashed horizontal lines are located at
the 50th percentiles.}
\label{figComparisonwithcox2}
\end{figure}

We compared our model with the feature-based time-dependent Cox
proportional hazard model (Cox PHM). We expect that the hazard rate
from our model and from the Cox PHM should be of similar accuracy;
however, if our assumption is true that the hazard rate can be split
into an external and an internal vulnerability state, the Cox PHM
should not be able to predict failures in advance as accurately as our
model, which uses the internal state to make decisions rather than the
total hazard.
The Cox PHM tends to be very sensitive to the covariate values at the
previous time step, making its hazard rate fluctuate and leading to
possible problems with decision-making. We used a time-dependent Cox
PHM with a Weibull-based baseline hazard function, trained on the same
training sets as our model. The Weibull distribution is the most
commonly used distribution in reliability and degradation analysis, and
is often used with the Cox PHM [\citet{Boutros20112102}]. We
repeated the experiments on all 5 splits of data, tuning the
regularization constant to 0.1 through cross-validation. To compare our
model with the Cox PHM, we calculated the hazard rank percentiles at
the failure point for both our model and for the Cox PHM for all five
test sets. We then counted the number of times in each test set that
our model gives a higher hazard rank, and then performed the sign test
to assess whether our model significantly outperformed the Cox PHM,
which it did ($p$-value${}={}$0.0480). The individual hazard ranks of each
lifetime using both models are shown in Figure~\ref{figComparisonwithcox2}(a). In Figure~\ref{figComparisonwithcox2}(b),
we show the box plots of rank percentile for our model and the Cox PHM.
These figures show that our method outperformed the Cox PHM for most
lifetimes and, in particular, that the vulnerability levels of failed
turbines were higher in our model than for the Cox~PHM.

%

\subsection{Cost-benefit analysis and decision-making}
As discussed in Section \ref{seccostbenefit}, the value of these
techniques lies in their power for maintenance decisions. We considered
cost function $C_d(\xi)$, shown in Figure~\ref{figsamplecdq}. We let
$c_1$ and $c_2$ denote the unit cost of a late 
and early warning, respectively. 
We assumed that early warnings are preferred over late warnings, that
is, $c_1\geq c_2$.
We applied the above model on all five training sets to find the
optimal threshold of $d = 5$ days and then calculated the associated
cost on the test sets.
The summary of results is given in Table~\ref{tabsummaryofwarningcost}.
We have also reported the total cost associated with warning at the
failure point, the cost associated with the Cox PHM, and the mean and
standard deviation (sd) of each model over the five folds. We repeated
this experiment on three different combinations of $c_1$ and $c_2$. As
$c_1$ was increased, the costs for all models also increased. Table~\ref{tabsummaryofwarningcost} indicates that our model performs slightly
better (across these combinations of $c_1$ and $c_2$) than the Cox PHM
with respect to cost; this is in addition to its distinct benefit of
being more interpretable. As expected, both models perform
substantially better than warning at failure.
%
\begin{table}
\tabcolsep=0pt
\caption{Summary of results (total cost) for the warning generation
process given our model and the Cox PHM for three combinations of $c_1$ and $c_2$}\label{tabsummaryofwarningcost}
\begin{tabular*}{\tablewidth}{@{\extracolsep{\fill}}@{}lcccccc@{}}
\hline
& \multicolumn{2}{c}{$\bolds{c_1=c_2}$}               & \multicolumn{2}{c}{$\bolds{c_1=5c_2}$}               & \multicolumn{2}{c@{}}{$\bolds{c_1=10c_2}$}\\
& \multicolumn{2}{c}{\textbf{cost of failure}}        & \multicolumn{2}{c}{\textbf{cost of failure}}         & \multicolumn{2}{c@{}}{\textbf{cost of failure}}\\
& \multicolumn{2}{c}{\textbf{warning${}\bolds{=95}$}} & \multicolumn{2}{c}{\textbf{warning${}\bolds{=475}$}} & \multicolumn{2}{c@{}}{\textbf{warning${}\bolds{=950}$}}
\\[-6pt]
& \multicolumn{2}{c}{\hrulefill} & \multicolumn{2}{c}{\hrulefill} & \multicolumn{2}{c@{}}{\hrulefill}
\\
\textbf{Test set} & \textbf{Our} & \textbf{Cox} & \textbf{Our} & \textbf{Cox} & \textbf{Our} & \textbf{Cox}\\
\textbf{No.} & \textbf{model} & \textbf{PHM} & \textbf{model} & \textbf{PHM} & \textbf{model} & \textbf{PHM}
\\
\hline
1 & 89 & 99 & 291 & 306 & 421 & 425 \\
2 & 95 & 91 & 180 & 210 & 255 & 198 \\
3 & 92 & 87 & 193 & 223 & 215 & 198 \\
4 & 92 & 95 & 364 & 393 & 404 & 484 \\
5 & 95 & 98 & 197 & 209 & 223 & 221
\\[3pt]
{Mean (sd)} & 92.6 (2.51) & 94.00 (5.0) & 245.0 (79.8) & 268.2 (80.5)& 303.60 (100.7)& 305.20 (138.2)\\
\hline
\end{tabular*}
\end{table}

\begin{figure}[b]

\includegraphics{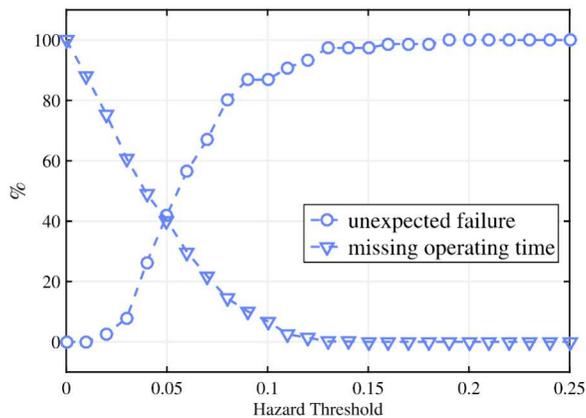}

\caption{Trade-off between the percentage of missing operating time and
the percentage of unexpected failures. This is shown over various
hazard thresholds using data from one of the test sets. Here, the
percentage of missing operating time is the fraction of total potential
operating time when the turbine does not operate due to early warnings.
The percentage of unexpected failures is the fraction of replacements
that happen as a result of late warnings. That is, it is the number of
failures that happen while the turbine is operating divided by the
total number of replacements. This trade-off is shown for the 19
lifetimes in one test set.}\label{figCost-BenefitAnalysis}
\end{figure}

Figure~\ref{figCost-BenefitAnalysis} illustrates another mechanism for
making decisions. It shows the trade-off between the percentage of
missing operating time and the percentage of unexpected failures. Here,
the percentage of missing operating time is the fraction of total
potential operating time when the turbine does not operate due to early
warnings. The percentage of unexpected failures is the fraction of
replacements that happen as a result of late warnings. That is, it is
the number of failures that happen while the turbine is operating
divided by the total number of replacements. This trade-off is shown
for the 19 lifetimes in one test set. A figure like this can be used to
determine the desired cost-benefit trade-off between early and late
warnings. Using the training data, we can then find the corresponding
threshold for the hazard rate to generate warnings, which depends on
the cost of failure replacements and the cost of nonfailure
replacements. One might choose the threshold on the hazard rate for
which the long-run average unit cost of the system is minimized.
%

\section{Numerical experiments}\label{secnumericalexp}
In this section we provide a set of numerical experiments, including a
simulation study and comparisons with previous models.

%
\begin{figure}

\includegraphics{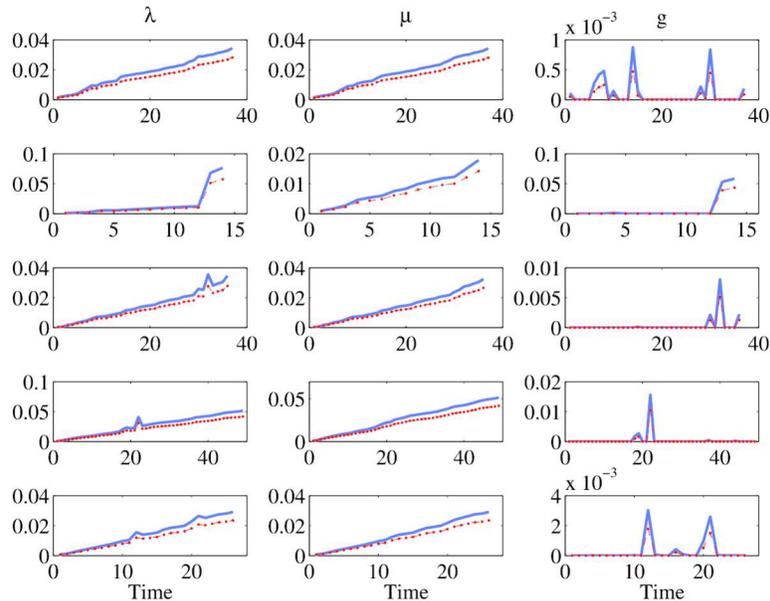}

\caption{True (solid lines) overall hazard rates ($\lambda$), internal
($\mu$) and external ($g$) hazard curves for 5 lifetimes (rows 1--5) and
their estimates (dashed lines) based on parameter estimation using
$N=50$ sample lifetimes.}\label{figsamplefeaturevaluessimulation}
\end{figure}

\subsection{Simulation study}\label{secsimulation}
In this section we demonstrate through simulation experiments (1) the
motivation of this work and (2) the empirical consistency of the
parameter estimation method for recovering true parameters. In our
simulation, a single feature is used as the observable signal over time
(we might consider this variable as representing the external
temperature near a wind farm). The rest of the parameters associated
with $\mu$ and $g$ used in this example are as follows:
$\beta_0=-7$, $\beta_1=0.5$, $\alpha_0=-14$, and $\alpha_1=5$. {These
values were chosen so that the parameters $\mu$ and $g$ were different,
but both would be on approximately the same scale}. These four
parameters specify the internal and external effects. Therefore, the
overall hazard rate for the $i$th lifetime at time $t$ is
\[
\lambda\bigl(t\mid x_i(1),\ldots,x_i(t)\bigr)= \biggl[
\sum_{\ell\in\{1, 2,3,\ldots, t\} }\exp\bigl(-7+ 0.5 x_i(\ell)
\bigr) \biggr]+\exp\bigl(-14 + 5 x_i(t)\bigr),
\]
where each $x_i(t)\sim\mathcal{N}(0,1)$. To generate a survival time
$T_i$ for each lifetime $i$ given the covariate measurements, we first
draw a random number $v$ from $\mathcal{U}(0,1)$. The failure time is
the time point at which the conditional survival function [which can be
computed at time $t$ as $\exp(-\int_0^t \lambda(u\mid x_i(1),\ldots
,\break x_i(u)) \,du )$] equals $v$. We then used $N$ (50, 100, 200, 400, 800)
simulated trajectories to estimate the parameters of the model. In
Figure~\ref{figsamplefeaturevaluessimulation}, the true internal and
external terms, and the overall hazard rates are shown as solid lines.
Their estimated values based on parameter estimation with $N= 50$
sample lifetimes are shown as dashed lines. Figure~\ref{figsamplefeaturevaluessimulation} shows that the model was able to
approximately capture both $\mu$ and $g$ terms. In this particular
experiment, the model slightly underestimated $\beta_0$ and $\beta_1$,
leading to a small bias in estimates for $\mu$ and $\lambda$ in all of
the lifetimes.
The discrepancy between the actual and the estimated rate increases as
a function of time (due to the cumulative nature of the hazard rate).

\begin{figure}

\includegraphics{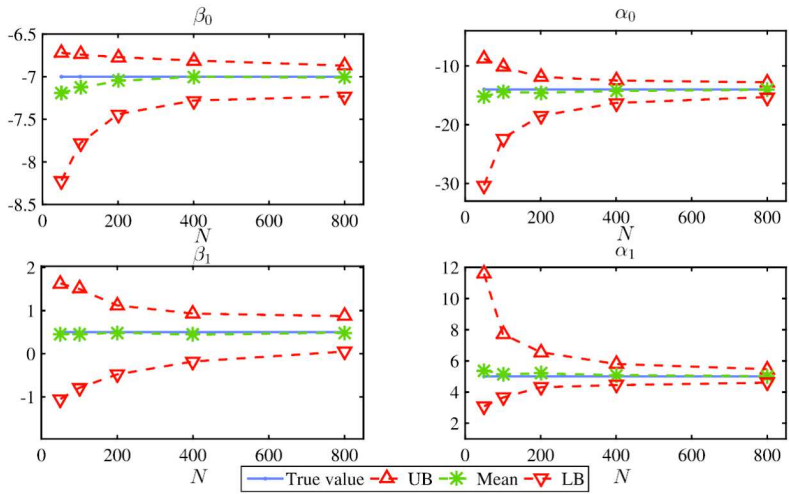}

\caption{The true value of each of the 4 model parameters $\alpha_0$,
$\alpha_1$, $\beta_0$ and $\beta_1$ (solid blue lines in the middle),
the corresponding mean estimate based on 100 runs (middle dashed lines
in blue) and the 95$\%$ prediction interval (the outer two dashed lines
in red), as a function of the number of sample lifetimes $N$ used for
estimation.}\vspace*{-3pt}
\label{figsampleinference}
\end{figure}

To evaluate the efficiency of the parameter estimation method, we
sampled lifetimes and used our method to recover the true parameters.
The simulation was repeated for $N ={}$50, 100, 200, 400 and 800
lifetimes to assess the convergence rate to the underlying true values.
To assess variance, the experiment was performed for 100 simulation
runs for each choice of $N$. We used the squared error between the
simulated and true parameter values to evaluate the estimation results.
In Figure~\ref{figsampleinference}, the mean estimate from the 100 runs
and its 95$\%$ prediction interval (the upper and lower bounds are
denoted by UB and LB, resp.) are shown for different values of
$N$ and the parameters of the model ($\alpha_0,\alpha_1,\beta_0,\beta
_1$). Figure~\ref{figsampleinference} illustrates that the prediction
intervals narrow and the estimated values converge to the true values
as $N$ increases.
Table~\ref{tablesim} in Appendix~\ref{app3} presents numerical values
for the\vadjust{\goodbreak} mean, the standard deviation and the mean squared error of
estimation over the 100 simulation runs for the various choices of $N$.
%

%
\begin{figure}
\includegraphics{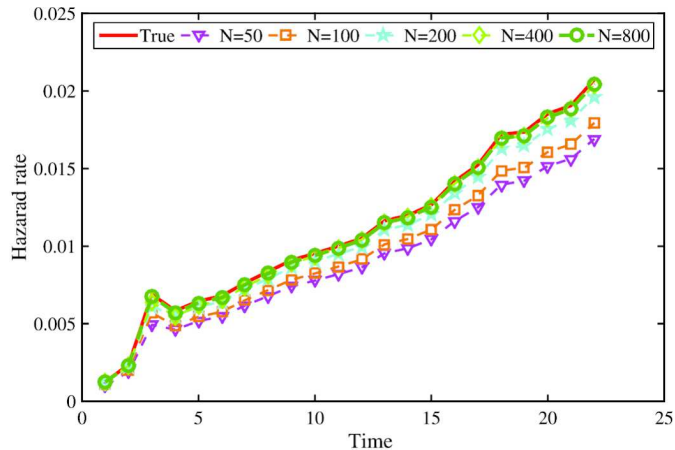}

\caption{Estimated results for $\lambda=\mu+g$ for one simulated
lifetime (dashed lines). The average percent errors for $N=50$, 100,
200, 400 and 800 are 19.06\%, 13.97\%, 5.59\%, 2.42\% and 1.40\%,
respectively. This figure shows that the estimation error decreases as
$N$ increases.}\vspace*{-3pt}\label{figsamplehazards}
\end{figure}

To demonstrate the quality of the estimation for the overall hazard
rate, we show the result of estimating $\lambda$ (overall hazard rate)
using $N$ simulated lifetimes for one randomly chosen lifetime and
various $N$ in Figure~\ref{figsamplehazards}. We can observe that the\vadjust{\goodbreak}
estimated hazard rates converge to the actual hazard rate as the number
of training data used for estimation increases. However, the
discrepancy between the actual and the estimated rate goes up as a
function of time (due to the cumulative nature of the hazard rate), and
this is particularly pronounced for the smaller values of~$N$.

\subsection{Comparison with existing models}
In this subsection we illustrate possible benefits of our model as
compared to the models in each of the three categories described in
Section~\ref{secrelatedwork}. The purpose is not to show that our model
outperforms previous methods with respect to the estimation of the
hazard rate; here, we aim to show that our model possesses the
complexity to reproduce behavior generated by existing approaches. We
will also show that our model has some additional benefits, such as
fewer assumptions and parameters.

\subsubsection{Comparison with models with degradation signals}
\begin{figure}[t]

\includegraphics{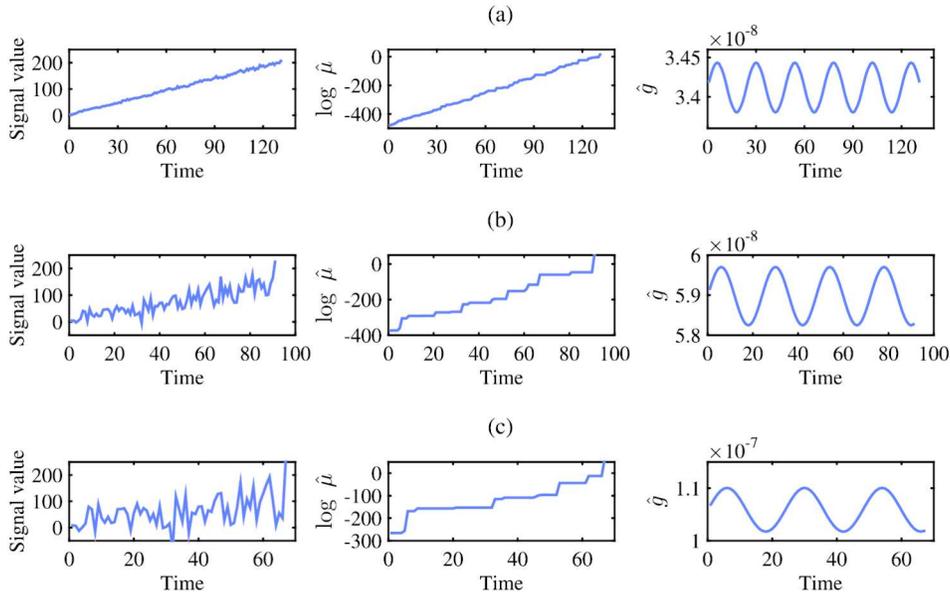}

\caption{Three lifetimes of degradation signals (column 1) generated
from the model of \citet{Bian2013}, and estimated $\log\hat{\mu}$
(column 2) and $\hat{g}$ from our model (column~3).
\textup{(a)}~Degradation signal, (1) $\gamma=1$;
\textup{(b)}~degradation signal, (2) $\gamma=5$;
\textup{(c)}~degradation signal, (3) $\gamma=10$.}\vspace*{3pt}\label{figcompdegradation1}
\end{figure}

We first show that our model can be easily made to reproduce behaviors
of other models that employ explicit degradation signals for
degradation modeling. We compare our model with one of the recent
papers in this category [\citet{Bian2013}], which employs
historical and real-time signals related to environmental conditions,
as well as an observable degradation signal representing the underlying
degradation process. The model of the degradation signal denoted by
$s_i(t)$ corresponding to the $i$th unit is expressed as
\begin{equation}
s_i(t)=\int_0^t\bigl[
\alpha_i+\beta_i\omega(v)\bigr]\,dv+\gamma_iB(t),
\end{equation}
where $\alpha, \beta$ and $\gamma$ are parameters of the model with
normal prior distributions, $\omega(t)$ is a deterministic
environmental condition that evolves according to a sine function:
$\omega(t)=2+\sin(\frac{\pi t}{12} )$, and $B(t)$ is a standard
Brownian Motion process. Unlike the assumptions for our model,
\citet{Bian2013} assumed (i) a predefined formula for the
degradation signal itself, (ii) a predefined threshold for failure
(predict failure when $s$ exceeds a predefined value), (iii) a known
time-dependent distribution with a sine function for the environmental
condition, and (iv) a Brownian noise distribution, and normal priors on
all other parameters. We simulated 1000 signals (500 for training and
500 for testing) from their model, and then used the simulated
degradation signals [$s_i(t), 1 \leq i \leq500$] and the environmental
observations to train our model. In Figure~\ref{figcompdegradation1},
%
%
%

\begin{figure}

\includegraphics{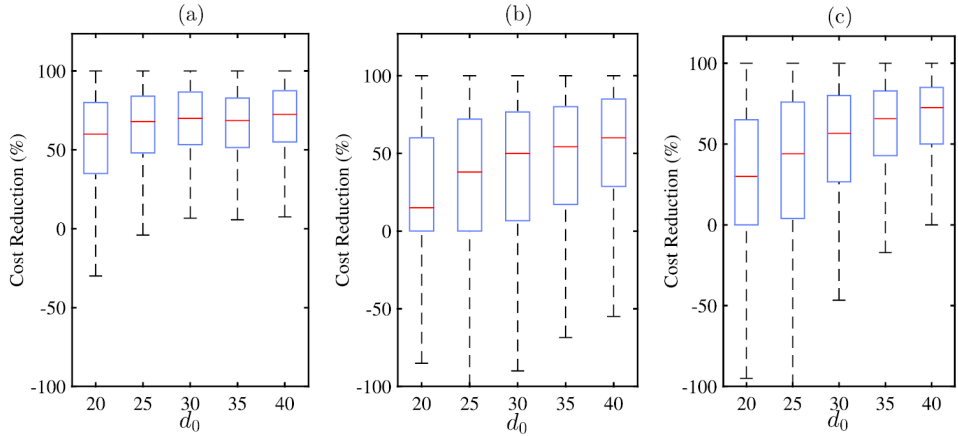}

\caption{Box plots for the cost reduction (\%) using our model for
three levels of noise ($\gamma$) and five levels of $d_0$.
\textup{(a)}~$\gamma=1$,
\textup{(b)}~$\gamma=5$,
\textup{(c)}~$\gamma=10$.}\vspace*{3pt}\label{figcompdegradation2}
\end{figure} %
three sample degradation signals and their corresponding estimates of
$\log(\hat{\mu})$ and $\hat{g}$ using our model are presented. Our
model managed to decompose the hazard rate perfectly into a monotonic
latent degradation function and a transient vulnerability rate---without needing to make the assumptions of \citet{Bian2013}
for the form of signal $s_i$. To assess whether the trained model is
useful for decision-making,
we used five different levels of leading time $d$ (assuming $c_1=c_2$),
and calculated the warning generation\vadjust{\goodbreak} cost reduction percentage using
our model with respect to warning at failure. From Figure~\ref{figcompdegradation2}
we can observe that the cost reduction distribution mostly (though not
always) takes positive values; in particular, the medians of the cost
reductions are\vadjust{\goodbreak} always positive and become larger as $\gamma$ decreases
(i.e., when there is less noise). This figure shows that, on
average, using the latent degradation state of our model to choose
warning times actually achieves reasonable performance with respect to
the warning generation process, despite the data being generated from
the model of \citet{Bian2013}.

\begin{figure}

\includegraphics{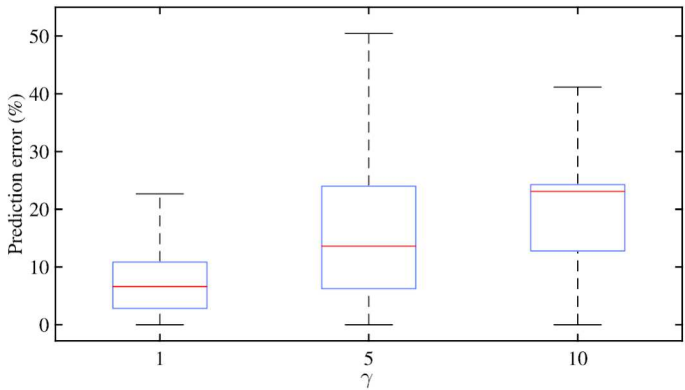}

\caption{Results (prediction errors) of remaining useful life
prediction as a function of the level of noise $\gamma$. These were
evaluated at the 75th percentile of the true lifetime (given that the
true remaining useful life is known).}\label{figrulprediction}
\end{figure}

\citet{Bian2013} aimed to predict the remaining useful life, so we
used our model to do the same. We evaluated predictions at the time of
the 75th percentile of the true lifetime. We first calculated $d$ as
the remaining life at the 75th percentile of the lifetime (we
calculated this for each simulated signal). Then, using the
corresponding threshold value of $d$ (called $\gamma_{d}$), we
calculated the warning time $(R_{i,\gamma_{d}})$ as explained in
Section~\ref{seccostbenefit}. This means our estimated lifetime is
$R_{i,\gamma_{d}}+d$. Then, we computed the prediction error using the
relative percentage difference, in the same way as \citet
{Bian2013}, which is
\[
\mbox{Prediction error}= 100 \times\frac{\llvert\mbox{Actual lifetime---Estimated lifetime}\rrvert }{\mbox{Actual lifetime}}.
\]
The box plots in Figure~\ref{figrulprediction} show that our model
provides remaining useful life predictions that are approximately as
accurate as those of Bian and Gebraeel's
[see Figure~3 in \citet{Bian2013}]. These results indicate that
our model could potentially represent a degradation model as complex as
that given by \citet{Bian2013}, using fewer parameters and without
making heavy distributional assumptions on the parameters or the
structure of the degradation signal or environmental conditions. Our
model also has the advantage that it does not require a predetermined
failure threshold, making the implementation of our model easier in
real-world problems.
%

\subsubsection{Comparison with hazard models with time-varying covariates}
We performed a set of experiments on prognostic turbofan engine data
from the publicly available NASA Prognostic Data Repository [\citet
{DeCastro2008,saxena2008}]. The data set, FD001, includes
run-to-failure time series signals that were collected from a dynamic
simulation process for 100 engines. These represent data from a modern
dual-spool, high-bypass ratio turbofan engine, which has been the focus
of many controls and diagnostics/health management studies over the
past few years
[see \citet{saxena2008} and references therein]. To generate
condition monitoring features, NASA developed a comprehensive logical
order of events that is similar to that of real engines. The engine
operates normally at the start of each time series and the fault grows
in magnitude until the system fails. Each record is a 24-element vector
that is a run-to-failure lifetime corresponding to a given operation
cycle. The vector consists of three values for the operational settings
and 21 values for engine performance measurements, which are averaged
over three cycles. We chose 3 cycles because it was computationally
more efficient to perform repeated experiments. {We did a small-scale
study with averaging removed, and we also considered averaging over 2
cycles. In both cases, the results were the same as averaging over 3
cycles.} All failures are caused by HPC (High-Pressure Compressor)
degradation. We randomly divided our data set into five folds (subsets)
of equal size (20 engines per fold). We trained the model with 4 folds
and used the last fold for testing, where 20 engines from the training
set were used as a validation set to fit the regularization constant.
This process was repeated 5 times so that all folds were used for testing.

%
\begin{table}
\tabcolsep=0pt
\caption{Summary of the hazard rank percentile on the turbofan data set
\textup{(TS1--TS5)} given by our model and the Cox PHM calculated at 1 and 10
cycles before failure}\label{tabhazardranknasa}
\begin{tabular*}{\tablewidth}{@{\extracolsep{\fill}}@{}lcccc@{\hspace*{-6pt}}}
\hline
& \multicolumn{2}{c}{\textbf{1 cycle before failure}} & \multicolumn{2}{c@{}}{\textbf{10 cycles before failure}}\\[-6pt]
& \multicolumn{2}{c}{\hrulefill} & \multicolumn{2}{c@{}}{\hrulefill}\\
\textbf{Test set No.}  & \textbf{Our model} & \textbf{Cox PHM} & \textbf{Our model} & \textbf{Cox PHM}\\
\hline
TS1 & 99.4 & 100.00 & 91.0 & 86.4 \\
TS2 & 97.9 & 100.0 & 79.8 & 86.1 \\
TS3 & 99.3 & \phantom{0}99.7 & 74.0 & 78.9 \\
TS4 & 98.5 & \phantom{0}99.3& 84.8 & 85.9 \\
TS5 & 99.5 & 100.0 & 90.0 & 92.8
\\[3pt]
Mean (sd) over 5 folds & 98.9 (0.6) & 99.8 (0.3) & 84 (7.1) & 86 (4.9)\\
\hline
\end{tabular*}
\end{table}

An interesting observation made from the trained models is that the
latent state hazard rate ($\mu$) dominates the transient hazard term
($g$), which means that failures are mainly the result of soft
degradation. This is consistent with the procedure by which these data
were generated (i.e., all failures are due to soft degradation) as
described by \citet{saxena2008}. The trained models suggest that
the hazard rate is very low during the early life of the engines but
increases significantly during the last 10\% of the lifetimes. We have
also observed that the hazard rate of each engine has some significant
jumps during its lifetime, which can potentially be an indication of
different damage levels.
We present an assessment of the failure prediction ability of our
model. We calculated the hazard rank percentile at the failure points
and 10 cycles before failure for all engines in the 5 test sets.
Results reported in Table~\ref{tabhazardranknasa} show that our method
performed well on all test sets, that is, at almost all cases the
hazard rank at the failure point and the hazard rank at 10 cycles
before the failure point were higher than those of other engines with
longer lifetimes. Our model and the Cox PHM performed comparably on all
test sets.

\begin{figure}[b]

\includegraphics{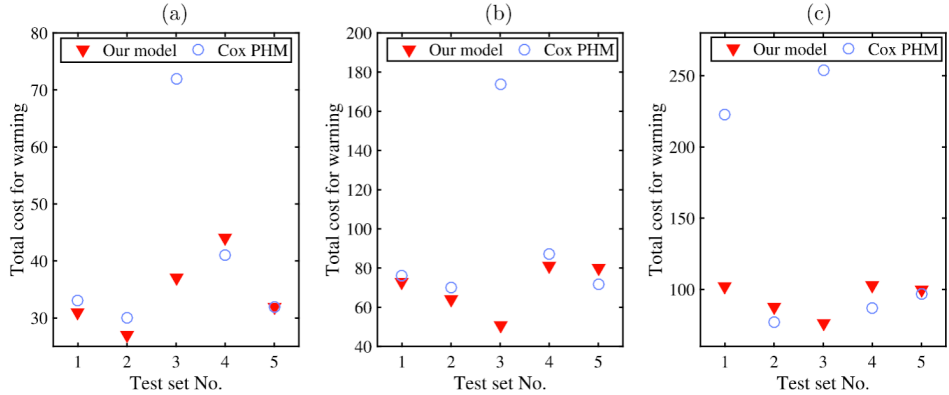}

\caption{Total cost for warning generation on the turbofan data set
\textup{(TS1--TS5)} given by our model and the Cox PHM for three combinations of
$c_1$ and $c_2$. The mean and standard deviation (sd) of the total cost
for our model for the three cases of $c_1=c_2$, $c_1=5c_2$ and
$c_1=10c_2$ are 34.20 (6.5), 69.80 (12.5) and 93.8 (11.6),
respectively. The mean and standard deviation (sd) of the total cost in
the Cox PHM model for the three cases of $c_1=c_2$, $c_1=5c_2$ and
$c_1=10c_2$ are 41.6 (17.5), 95.80 (44.2) and 147.6 (84),
respectively.
\textup{(a)}~$c_1=c_2$,
\textup{(b)}~$c_1=5c_2$,
\textup{(c)}~$c_1=10c_2$.}\label{figtable4}
\end{figure}

To demonstrate the potential of using our model for warning generation,
we applied the model to all 5 splits of training data to find the
optimal threshold when $d = 5$ cycles and then calculated the
associated cost on the test sets. The summary\vadjust{\goodbreak} of results given in
Table~\ref{tabcostfornasa} and Figure~\ref{figtable4} indicates that
our model, as expected, performs better than warning at failure.
Compared to the Cox PHM, although the differences between the two
models may not be significantly different, our model generally tends to
result in a lower average cost and standard deviation than the Cox PHM.

%
%
\begin{table}
\tabcolsep=0pt
\caption{Summary of results (total cost for warning generation on
turbofan data set) given by our model and the Cox PHM for three
combinations of $c_1$ and $c_2$}\label{tabcostfornasa}
\begin{tabular*}{\tablewidth}{@{\extracolsep{\fill}}@{}@{}lcccccc@{}}
\hline
& \multicolumn{2}{c}{$\bolds{c_1=c_2}$}               & \multicolumn{2}{c}{$\bolds{c_1=5c_2}$}               & \multicolumn{2}{c@{}}{$\bolds{c_1=10c_2}$}\\
& \multicolumn{2}{c}{\textbf{cost of failure}}        & \multicolumn{2}{c}{\textbf{cost of failure}}         & \multicolumn{2}{c@{}}{\textbf{cost of failure}}\\
             & \multicolumn{2}{c}{\textbf{warning${}\bolds{=100}$}}& \multicolumn{2}{c}{\textbf{warning${}\bolds{=500}$}} & \multicolumn{2}{c@{}}{\textbf{warning${}\bolds{=1000}$}}
\\[-6pt]
& \multicolumn{2}{c}{\hrulefill} & \multicolumn{2}{c}{\hrulefill} & \multicolumn{2}{c@{}}{\hrulefill}
\\
\textbf{Set} & \textbf{Our} & \textbf{Cox} & \textbf{Our} & \textbf{Cox} & \textbf{Our} & \textbf{Cox}\\
\textbf{No.} & \textbf{model} & \textbf{PHM} & \textbf{model} & \textbf{PHM} & \textbf{model} & \textbf{PHM}
\\
\hline
TS1 & 31 & 33 & 73 & \phantom{0}76 & 102 & 223 \\
TS2 & 27 & 30 & 64 & \phantom{0}70 & \phantom{0}88 & \phantom{0}77 \\
TS3 & 37 & 72 & 51 & 174 & \phantom{0}76 & 254 \\
TS4 & 44 & 41 & 81 & \phantom{0}87 & 103 & \phantom{0}87 \\
TS5 & 32 & 32 & 80 & \phantom{0}72 & 100 & \phantom{0}97
\\[3pt]
Mean (sd)  \\
\quad over 5 \\
\quad folds & 34.20 (6.5) & 41.60 (17.5) & 69.80 (12.5) & 95.80 (44.2) & 93.80 (11.6) & 147.60 (84.0)\\
\hline
\end{tabular*}
\end{table}

%
%

%
\begin{figure}

\includegraphics{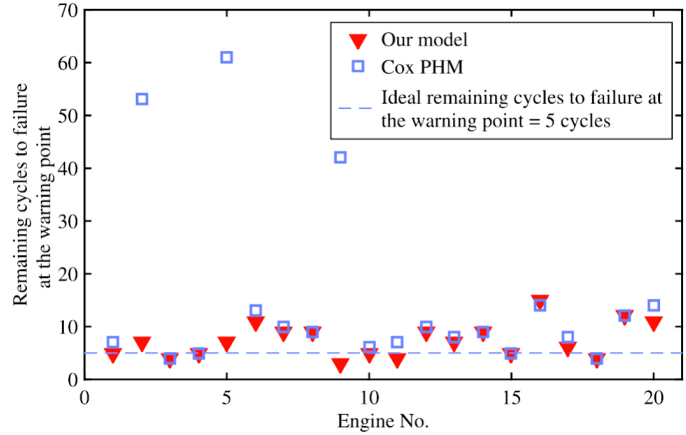}

\caption{The remaining cycles to failure calculated at warning points
from our model and the Cox PHM. Ideally, the warning should be issued
as close as possible to 5 cycles before failure (dashed line). The
suggested warning times for Engines 2, 5 and 9 are too early using the
Cox PHM. For instance, for Engine 2, the Cox PHM model issued a warning
over 53 cycles too early, whereas our model issued a warning that was
near the ideal of 5 cycles before failure.
}\label{figwarningtimeexample}
\end{figure}

\begin{figure}[b]

\includegraphics{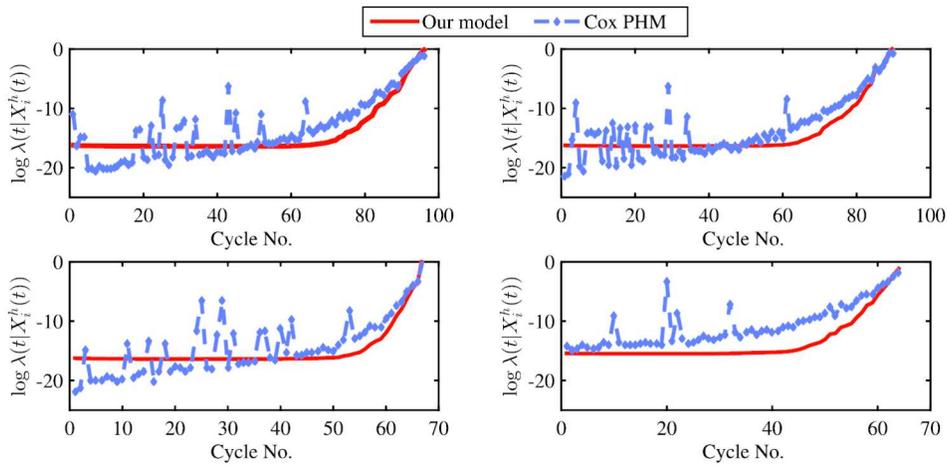}

\caption{Sample estimated log of the hazard rates from our model (solid
lines) and the Cox PHM (dashed lines).}
\label{figourmodelandcoxexample}
\end{figure}

The remaining cycles to failure calculated at the suggested warning
times given by our model and the Cox PHM for the 20 engines in the
first training set are shown in Figure~\ref{figwarningtimeexample}. It
can be seen from this figure that the remaining cycles to failure at
the warning point given by our model are closer to the desired one,
which is 5 cycles. The Cox PHM had some predictions that were very
poor, where the warning time was much too early. This is due to
problems with robustness of the Cox PHM that our model does not
generally have due to its natural regularization resulting from the use
of the internal state. As shown in
Figure~\ref{figourmodelandcoxexample}, our model is more robust with
respect to changes in covariates, which leads to better decision-making.

%
%

%
\begin{figure}

\includegraphics{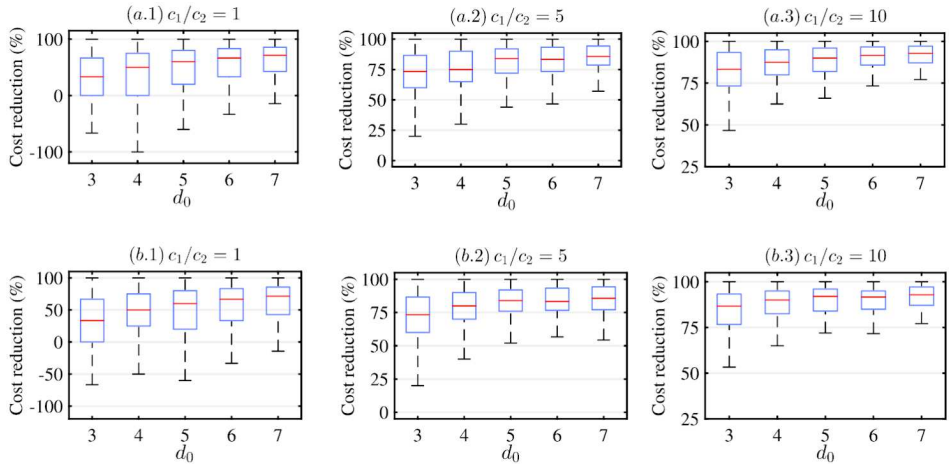}

\caption{Cost reduction comparison between \textup{(a.1)--(a.3)} our model and
\textup{(b.1)--(b.3)} hidden multistate models for three values of $c_1/c_2$. The
upper plots are similar to their corresponding lower plots. The same
trends are visible within each upper plot and lower plot.}
\label{fighmm}
\end{figure}
%

\subsubsection{Comparison with partially-observed degradation
models}\label{22222}
We compared our model with one of the recent models in the literature
[\citet{Ghasemi201045}], which used a discrete, hidden multistate
stochastic process for degradation modeling. The degradation process
$Z(t)$ is assumed to be a three-state Markov process with a transition
matrix $\mathbf{P}$. The states are only indirectly observable through
condition monitoring. The output of condition monitoring at time $t$,
denoted by $y_t$, is one of five possible values, and $y_t$ is
stochastically related to the actual level of degradation. This
stochastic relationship is represented by matrix $\mathbf
{Q}=[q_{j,i}]$, where $q_{j,i}$ is the probability of getting the $i$th
output $(i \in\{1,2,\ldots,5\})$ while the system is in degradation
state $j$, $j\in\{1,2,3\}$. From the degradation state, failures are
generated according to a time-dependent proportional hazard model. The
hazard function is
\[
\lambda\bigl(t,Z(t)\bigr)=\frac{\zeta}{\eta} \biggl(\frac{t}{\eta}
\biggr)^{\zeta-1} \exp\bigl(\gamma Z(t)\bigr),
\]
where $\zeta$ and $\eta$ are the scale and shape parameters associated
with a Weibull baseline and $\gamma$ is the regression coefficient. The
parameters used in our analysis are given in Appendix~\ref{app4}. We
simulated 1000 lifetimes based on this model, 500 for training and 500
for testing. We used $\log(t)$ and $y_t$ as our covariates.
We compared the cost reduction of using our model with that of the
model of \citet{Ghasemi201045} over multiple combinations of $d$
and $c_1/c_2$. Figure~\ref{fighmm} shows that (i) as $c_1/c_2$
increases, there is more cost reduction for both models and (ii) our
model has very similar performance to that of \citet{Ghasemi201045}.
Our model has the benefits that it has fewer parameters, and training
our model is computationally less expensive. Our model has only 5
parameters $(\alpha_0,\alpha_1,\alpha_2,\beta_0,\beta_1,\beta_2)$, but
the hidden Markov process described here has 27 parameters (9 for
$\mathbf{P}_{3\times3}$, 15 for $\mathbf{Q}_{3 \times5}$ and 3 for
$\eta$, $\zeta$ and $\gamma$).


\section{Concluding remarks}\label{secconclusion}
We presented a method for separating the latent internal hazard rate
from the temporary hazard due to external sources. We showed that the
method has some major advantages over the Cox proportional hazard
model, in that it can encode the full history of the turbine within the
estimated degradation state in a natural way that the Cox PHM cannot.
Further, because the latent degradation state is estimated, it can be
used for making maintenance decisions. There are many possible
extensions and uses for this model. Although we designed the model for
predictive maintenance at wind farms, it can be used for any type of
equipment failure prediction, health condition maintenance and in many
other application domains (e.g., healthcare). It is also possible that
additional prior knowledge is available about the influence of the
external factors on the latent state, which can be incorporated in an
extended version of the model. In cases where self-healing is possible
due to external factors, a third term (in addition to $\mu$ and $g$)
that reduces the hazard rate could be introduced. The latent state
hazard model has the benefit that it requires very few distributional
assumptions and can be trained in a computationally efficient way
through convex optimization.

\begin{appendix}
\section{Proof of Proposition \texorpdfstring{\protect\ref{11111}}{1}}\label{app1}
The proof relies on the fact that the composition of convex functions
is convex, as follows.

\begin{lemma}{Composition rule for convex functions.}\label
{lemmaconvexfunctions}
Suppose
\[
f(x) = h \bigl(\omega_1(x), \omega_2(x), \ldots,
\omega_k(x) \bigr),
\]
where $h: \mathbb{R}^k \to\mathbb{R}$ is convex, and
$\omega: \mathbb{R}^n \to\mathbb{R}$. The function f is convex if one
of the following holds:
\begin{itemize}
\item$h$ is nondecreasing in the $ith$ argument, and $\omega_i$ is convex.
\item$h$ is nonincreasing in the $ith$ argument, and $\omega_i$ is concave.
\end{itemize}
\end{lemma}

\begin{pf*}{Proof of Proposition \ref{11111}}
It is sufficient to prove that
\[
H_{i,j}(\bolds{\theta})=-\log\bigl( y_{i,j}+(1-2
y_{i,j})\exp\bigl( - \lambda\bigl( {j}\mid\mathbf{x}_i^h
( {j} ) \bigr) \bigr) \bigr)
\]
is convex in $\bolds{\theta}$ for all $(i,j)$, based on the fact that the
sum of convex functions is also convex.
Now, if $y_{i,j}=0$, then
\[
H_{i,j}(\bolds{\theta})=\lambda\bigl( {j}\mid\mathbf{x}_i^h
( {j} ) \bigr),
\]
which is convex in $\bolds{\theta}$ due to the fact that the sum of two
convex functions $\mu(t\mid\mathbf{x}_i^h(t))$ and $g(t\mid\mathbf{x}_i(t))$
is convex. Recall that $g(t\mid\mathbf{x}_i(t))$ is the exponential of an
affine function and $\mu(t\mid\mathbf{x}_i^h(t))$ is a sum of exponentials
of affine functions, which are both convex in $\bolds{\theta}$. If
$y_{i,j}=1$, then
\[
H_{i,j}(\bolds{\theta})=-\log\bigl( 1 - \exp\bigl( - \lambda\bigl({j}
\mid\mathbf{x}_i^h ( {j} ) \bigr) \bigr) \bigr).
\]
We will invoke Lemma \ref{lemmaconvexfunctions}. Since $-\log(\cdot)$
is nonincreasing, we have that $-\log f(\cdot) $ is convex if $f$ is
concave and positive. That is, $H_{i,j}(\bolds{\theta})$ for $y_{i,j}=1$
is convex only if
$
( 1 - \exp( - \lambda({j}\mid\mathbf{x}_i^h
({j})) ) )
$
is concave and positive. Now, since the hazard rate is always
nonnegative by its definition, then $\lambda({j}\mid x^h_i(j) ) > 0$
and,thus, $\exp ( - \lambda( {j}\mid x^h_i(j) ) ) < 1 $, so
\[
\bigl( 1 - \exp\bigl( - \lambda\bigl( {j}\mid\mathbf{x}_i^h
({j} ) \bigr) \bigr) \bigr) > 0.
\]
Also, since $\lambda( {j}\mid\mathbf{x}_i^h ( {j} ) )$ is convex in $\bolds
{\theta}$ (i.e., the sum of two convex functions of $\mu$ and $g$ is
also convex), then
$ ( 1 - \exp( - \lambda( {j}\mid\mathbf{x}_i^h ) ) )\mbox{ is
concave in } \bolds{\theta}$.
We can conclude that for $y_{i,j}=1$, $H_{i,j}(\bolds{\theta})$ is convex
in~$\bolds{\theta}$. We now have that $H_{i,j}(\bolds{\theta})$ is convex
in~$\bolds{\theta}$ for all $(i,j)$. Also, note that the $\ell_2$
regularization term is strictly convex and, therefore,
\[
W_N(\bolds{\theta}\mid X)=-\log\mathcal{L}_N(\bolds{\theta}
\mid X)+ C_1 \llVert\bolds{\alpha}\rrVert^2_2
+ C_2 \llVert\bolds{\beta}\rrVert^2_2
\]
is strictly convex.\
\end{pf*}

\section{Asymptotic properties}\label{app2}
To our knowledge, no work on AMMHM has studied asymptotic properties
for a class of models that includes ours. Thus, we prove basic
asymptotic properties of the maximum likelihood estimators of the
parameters of our model here. There are regularity conditions for the
maximum likelihood estimate that can guarantee consistency and
asymptotic normality [\citet{lehmann1998}]. We verify these
regularity conditions for our model. Similar analysis was done in
\citet{Mamunur2009} for the logistic regression model. Let $X_i, 1
\leq i \leq N$ be i.i.d. random variables with a p.d.f. $f(\bolds{\theta},X_i)$
that depends on parameters $\bolds{\theta} \in\Theta\subseteq\mathbb
{R}^p$. The regularity conditions are as follows:
\begin{longlist}[(C6)]
\item[(C0)] The distributions $f(\bolds{\theta},X)$ of the
observations are distinct (otherwise, $\bolds{\theta}$ cannot be estimated
consistently).

\item[(C1)] The distributions $f(\bolds{\theta},X)$ have common support.

\item[(C2)] The random variables are $X=(X_1,\ldots, X_N)$, where the
$X_i$'s are i.i.d. with probability density $f(\bolds{\theta},X_i)$ with
respect to probability measure $\mu$.

\item[(C3)] There exists an open subset $\omega$ of $\Omega$ containing the true
parameter point $\bolds{\theta}^0$ such that for almost all $x$ the
density $f(\bolds{\theta},X)$ admits all third derivatives
$(\frac{\partial^3}{\partial{\theta_k}
\partial{\theta_j}\,\partial{\theta_z}}f(\bolds{\theta},X))$ for all $\bolds{\theta}\in\omega$.\vspace*{1pt}

\item[(C4)] The first and second logarithmic derivatives of $f(\bolds
{\theta},X)$ satisfy the equations
\begin{eqnarray}
\label{eqc41}
\mathbb{E} \biggl[ \frac{\partial}{\partial\theta_k}\log f(\bolds{\theta},X) \biggr] &=&0\qquad \forall k,
\\
%
I_{jk}(\bolds{\theta})&=&\mathbb{E} \biggl[
\frac{\partial}{\partial\theta_j} \log f(\bolds{\theta},X) \frac{\partial
}{\partial\theta_k} \log f(\bolds{\theta},X)
\biggr]
\nonumber\\[-8pt]\label{eqc42}  \\[-8pt]\nonumber
& =& \mathbb{E} \biggl[ - \frac{\partial^2}{\partial\theta
_j\,\partial\theta_k} \log f(\bolds{\theta},X) \biggr]\qquad
\forall j,k.
\end{eqnarray}

\item[(C5)] Since the $p\times p$ matrix $\mathbf{I}(\bolds{\theta})$
is a covariance matrix, it is positive semidefinite. We shall assume
that the $I_{j,k}(\bolds{\theta})'s$ are finite and that the matrix
$\mathbf{I}(\bolds{\theta})$ is positive definite for all $\bolds{\theta}$ in
$\omega$, and hence that the statistics
\[
\frac{\partial}{\partial_1} \log f(\bolds{\theta},X),\ldots, \frac{\partial
}{\partial_p} \log f(\bolds{
\theta},X)
\]
are affinely independent with probability 1.

\item[(C6)] Finally, we will assume that there exists function
$M_{k,m,z}$ such that
\[
\biggl\llvert\frac{\partial{{\partial}^3}}{\partial{\theta_k}\,\partial{\theta_m}
\,\partial{\theta_z}} \log f(\bolds{\theta},X)\biggr
\rrvert\leq
M_{k,m,z}(X)\qquad\mbox{for all } \bolds{\theta} \in\omega,
\]
where $m_{k,m,z}=\mathbb{E} [ M_{k,m,z}(X) ] < \infty $ for all $k,m,z$.

If the above assumptions are satisfied, the following theorem obtained
from \citet{lehmann1998} can be used for the asymptotic properties
of the maximum likelihood estimator.
\end{longlist}

%
\begin{theorem}\label{theoremcconditions}
Let $X_1,\ldots, X_N$ be i.i.d. each with a density $f(\bolds{\theta},X)$, with
bounded $\bolds{\theta}$ and $X$, which satisfies \textup{(C0)--(C6)}. Then with
probability tending to 1 as $N$ tends to infinity, there exist solutions
$\hat{{\bolds{\theta}}}_N=
(\hat{{{\theta}}}_{N,1},\ldots,\hat{{{\theta}}}_{N,p})$ of the
likelihood equations such that:
\begin{longlist}[(iii)]
\item[(i)] $\hat{{{\theta}}}_{N,j}$ is consistent for estimating
$\theta_j$,

\item[(ii)] $\sqrt{N} (\hat{{\bolds{\theta}}}_{N}-\bolds{\theta}^0)$ is
asymptotically normal with (vector) mean zero and covariance
matrix $[\mathbf{I}(\bolds{\theta})]^{-1}$, and

\item[(iii)] $\hat{{{\theta}}}_{N,j}$ is asymptotically efficient in
the sense that its variance attains the Cram\'er--Rao lower bound as $N$
goes to infinity.
\end{longlist}
\end{theorem}

The following theorem states that the regularity conditions are
satisfied in our model. Here, $f(\bolds{\theta},X)$ should be replaced
with the likelihood probability given in equation (\ref{eqlikelihoodmain}).

%
\begin{theorem}\label{theoregularity}
Let $X=(X_i, T_i), 1 \leq i \leq N$, be i.i.d. with the likelihood
function $\mathcal{L}_N(\bolds{\theta}\mid X)$ given in equation (\ref
{eqlikelihoodmain}) with independent parameter set $\bolds{\theta} \in
\Theta$, $\llvert \bolds{\alpha}\rrvert <M$, and $\llvert \bolds{\beta
}\rrvert <M$, where $M$ is a finite positive constant independent of
$N$. Then, the regularity conditions \textup{(C0)$,\ldots,$(C6)} are
satisfied for this model.
\end{theorem}
%

\begin{pf}
Each regularity condition is verified separately as follows:
\begin{longlist}[(C0)]
\item[(C0)] The condition (C0), also called
identifiability, refers to the fact that the true but unknown
parameters of the model ($\bolds{\theta}^0$) should be identified
(estimable). The parameters $\bolds{\theta}$ are identified if for any
parameter vector $\bolds{\theta}'$ $(\bolds{\theta}' \neq\bolds{\theta}^0)$,
for some $X$, $\mathcal{L}_N(\bolds{\theta}'\mid X)\neq L_N(\bolds{\theta
}^0,X)$. The log-likelihood of our model is an additive multi-index
model, which is a linear combination of nonlinear transformations of a
linear combination of explanatory variables. Since the parameter space
is convex and the regularized log-likelihood function is strictly
convex, then the solution of the maximum likelihood problem is unique
and, therefore,
the true parameter set $\bolds{\theta}^0$ is identified.

\item[(C1)] To show that all distributions $P_{\bolds{\theta}}$ have
common support, we can prove without loss of generality that the set
$A=\{x\mid\mathcal{L}_N(\bolds{\theta}\mid X)>0\}$ is independent of~$\bolds
{\theta}$. For the probability distribution of our model, since
$\llvert \alpha_i \rrvert < M$, $\llvert \beta_i \rrvert < M$, and
feature values are also bounded, then $\mathcal{L}_N(\bolds{\theta}\mid
X)$ is always greater than zero [i.e., the two elements of the
likelihood function given in equation (\ref{eqlikelihoodmain}) are
greater than zero regardless of $\bolds{\theta}$]. The only exception is
when all feature values are 0, which would make the likelihood 0
regardless of $\bolds{\theta}$.

\item[(C2)] This condition is one of the assumptions of the model
whereby unit lifetimes are assumed to be i.i.d. with probability
distribution generated from the hazard rate function given in
equations (\ref{eqhazardfunctionmain}). Thus, condition (C2) is
satisfied.

\item[(C3)] As the log-likelihood function is a linear combination
of other nonlinear functions for each $(i,j)$, it is sufficient to
prove that the third derivative exists for each
choice of $(i,j)$ with respect to $\bolds{\theta}$. It is easy to show
that the third derivative of each observation in $\mathcal{L}_N(\bolds
{\theta}\mid X)$ exists if the first, second and the third derivatives
of $\lambda (j\mid\mathbf{x}_i^h(j) )$ exist for all $(i,j)$. Since
$\lambda=\mu+g$, then we can show that condition~(C3) is met
if the first, second, and the third derivative of $\mu$ and $g$ exists
for all $\bolds{\theta} \in\omega$. Now, we can show that
\begin{eqnarray}
\label{eqthirdg1} \cases{
\displaystyle\frac{\partial\lambda({j}\mid\mathbf
{x}_i^h({j}))}{\partial\alpha_k}=x_{i,k}(j) \exp\bigl(\bolds{
\alpha}^\top\mathbf{x}_i(j)\bigr),
\vspace*{3pt}\cr
\displaystyle \frac{\partial^2\lambda({j}\mid\mathbf{x}_i^h({j}))}{\partial\alpha
_k\,\partial\alpha_m}=x_{i,k}(j) x_{i,m}(j) \exp\bigl(\bolds{\alpha}^\top
\mathbf{x}_i(j)\bigr),
\vspace*{3pt}\cr
\displaystyle\frac{\partial^3\lambda({j}\mid\mathbf{x}_i^h({j}))}{\partial\alpha
_k\,\partial\alpha_m \,\partial\alpha_z}= x_{i,k}(j)
x_{i,m}(j) x_{i,z}(j) \exp\bigl(\bolds{\alpha}^\top
\mathbf{x}_i(j)\bigr),}
\\
%
\label{eqthirdg2}
\cases{ \displaystyle\frac{\partial\lambda({j}\mid\mathbf
{x}_i^h(j))}{\partial\beta_k}=\sum_{l=1}^{j}
x_{i,k}(l) \exp\bigl(\bolds{\beta}^\top\mathbf{x}_i(l)\bigr),
\vspace*{3pt}\cr
\displaystyle\frac{\partial^2\lambda({j}\mid\mathbf{x}_i^h(j))}{\partial\beta_k\,\partial
\beta_m}= \sum_{l=1}^{j}
x_{i,k}(l) x_{i,m} (l) \exp\bigl(\bolds{\beta}^\top
\mathbf{x}_i(l)\bigr),
\vspace*{3pt}\cr
\displaystyle\frac{\partial^3\lambda({j}\mid\mathbf{x}_i^h({j}))}{\partial\beta_k
\,\partial\beta_m \,\partial\beta_z}= \sum
_{l=1}^{j} x_{i,k}(l) x_{i,m}(l)
x_{i,z}(l) \exp\bigl(\bolds{\beta}^\top\mathbf{x}_i(l)
\bigr).}
\end{eqnarray}
Since all of the above derivatives exist for all $\bolds{\theta}$ (given
that $\bolds{\theta}$ and $X$ are bounded), then (C3) is
satisfied for our model.

\item[(C4)] This condition shows two properties of the score
function and the Hessian matrix of the log-likelihood function. To
verify these relationships, we first rewrite the first derivative of
the log-likelihood function in terms of a sum over time $u=1,\ldots,\max
\{T_1,\ldots,T_N\}$ using equation \eqref{eqgradient} as
%
\begin{eqnarray}
&& \frac{\partial}{\partial\theta_k} \bigl(\log\mathcal{L}_N(\bolds{\theta }\mid X)\bigr)\nonumber
\\
&&\qquad =\sum_{u} \sum_{i}
-Y_i(u-1) (1-2y_{i,u}) \frac{\exp(-\lambda({u}\mid\mathbf{x}_i^h({u})) )}{
y_{i,u}+(1-2y_{i,u})\exp(-\lambda(u\mid\mathbf{x}^h_i({u})) )}
\\
&&\quad\qquad{}\times
\frac{\partial\lambda({u}\mid\mathbf{x}_i^h({u}))}{\partial\theta_k},\nonumber
\end{eqnarray}
where $u$'s are the time points and $Y_i(u-1)$ is a binary indicator
function being one only if unit $i$ has survived until time point
$u-1$. This allows the sum over $u$ to range over all time instead of
only until the failure times as in equation \eqref{eqgradient}. The
term $\mathbb{E} [\frac{\partial}{\partial\theta_k} (\log\mathcal
{L}_N(\bolds{\theta}\mid X))] $ can now be calculated as
%
\begin{eqnarray}
\label{eqexgradient}
&& \mathbb{E} \biggl[\frac{\partial}{\partial\theta
_k} \bigl(\log\mathcal{L}_N(\bolds{\theta}\mid X)\bigr) \biggr]\nonumber
\\
&&\qquad = \sum_{u} \sum_{i} \mathbb{E}
\\
&&\quad\qquad{}\times \underbrace{ \biggl[ -Y_i(u-1) \frac{(1-2y_{i,u})\exp
(-\lambda({u}\mid\mathbf{x}_i^h({u})) )}{
y_{i,u}+(1-2y_{i,u})\exp(-\lambda({j}\mid\mathbf{x}^h_i({u})) )}
\frac{\partial\lambda({u}\mid\mathbf{x}_i^h({u}))}{\partial\theta_k} \biggr
]}_{F(i,u,k)}.\nonumber
\end{eqnarray}
So, if $Y_i(u-1)$ is zero, the internal quantity is zero. Otherwise, we
can expand $y_{i,u}$ based on the two possibilities of (1) failure
within the next interval $(u-1,u)$ with the probability of
$ (1- \exp(-\lambda(u\mid\mathbf{x}_i^h(u))) )$,
and (2) survival within the next interval with the probability of
$ (\exp(-\lambda(u\mid\mathbf{x}_i^h(u))) )$. Considering the interior of
equation~\eqref{eqexgradient}, we have
%
\begin{eqnarray}
\label{eqproofC41}
&& \mathbb{E}\bigl(F(i,u,k)\bigr)\nonumber
\\
&&\qquad  =-Y_i(u-1)
\mathbb{E} \biggl[ \frac{\partial\lambda({u}\mid\mathbf
{x}_i^h({u}))}{\partial\theta_k} \times\bigl(\exp\bigl(-\lambda\bigl
(u\mid
\mathbf{x}_i^h(u)\bigr)\bigr) \bigr)
\nonumber\\[-8pt]\\[-8pt]\nonumber
&&\quad\qquad {} - \frac{\exp(-\lambda({u}\mid\mathbf{x}_i^h({u})) )}{
1-\exp(-\lambda({j}\mid\mathbf{x}^h_i({u})) )} \frac{\partial\lambda
({u}\mid\mathbf{x}_i^h({u}))}{\partial\theta_k} \times\bigl(1-\exp\bigl(-
\lambda\bigl(u\mid\mathbf{x}_i^h(u)\bigr)\bigr) \bigr)
\biggr]
\\
&&\qquad = - Y_i(u-1) \times0 = 0,\nonumber
\end{eqnarray}
which proves equation \eqref{eqc41} in (C4). We use a similar
approach to prove equation~(\ref{eqc42}).
The left-hand side of equation (\ref{eqc42}) can be calculated as
%
\begin{eqnarray}
\label{eqFi,u,k} &&\mathbb{E} \biggl[ \sum_{u} \sum
_{u'} \sum_{i} \sum
_{i'} F(i,u,k) \times F\bigl(i',u',m
\bigr) \biggr]\nonumber
\\
\nonumber
&&\qquad =\mathbb{E} \biggl[ \sum_{u} \sum
_{i} F(i,u,k) \times F(i,u,m) \biggr]
\\
&&\quad\qquad{} + \mathbb{E}
\biggl[ \sum_{u} \sum_{i}
\sum_{ i'\neq i} F(i,u,k) \times F\bigl(i',u,m
\bigr) \biggr]
\\
\nonumber
&&\quad\qquad{}+ \mathbb{E} \biggl[ \sum_{u} \sum
_{u' \neq u} \sum_{i} F(i,u,k)
\times F\bigl(i,u',m\bigr) \biggr]
\\
&&\quad\qquad{} + \mathbb{E} \biggl[ \sum
_{u} \sum_{u' \neq u} \sum
_{i} \sum_{i' \neq i} F(i,u,k)
\times F\bigl(i,u',m\bigr) \biggr].\nonumber
\end{eqnarray}
The first part of the above equation can be rewritten using equation
\eqref{eqexgradient} as
%
\begin{eqnarray}
\label{eqLHS1} &&\sum_{u} \sum
_{i} \mathbb{E} \bigl[ F(i,u,k) \times F(i,u,m) \bigr]\nonumber
\\
\nonumber
&&\qquad = \sum_{u} \sum
_{i} \mathbb{E} \biggl[Y_i(u-1)^2
\biggl( \frac{\exp (-\lambda({u}\mid\mathbf{x}_i^h({u})) )}{
y_{i,u}+(1-2y_{i,u})\exp(-\lambda({j}\mid\mathbf{x}^h_i({u})) )} \biggr
)^2
\\
&&\quad\qquad{}\times\frac{\partial\lambda({u}\mid\mathbf{x}_i^h({u}))}{\partial
\theta_k}
\frac{\partial\lambda({u}\mid\mathbf{x}_i^h({u}))}{\partial\theta_m} \biggr]\nonumber
\\
&&\qquad = \sum_{u} \sum
_{i} Y_i(u-1)^2 \frac{\partial\lambda({u}\mid\mathbf{x}_i^h({u}))}{
\partial\theta_k}
\frac{\partial\lambda({u}\mid\mathbf{x}_i^h({u}))}{\partial{\theta_m}}
\nonumber\\[-8pt]\\[-8pt]\nonumber
&&\quad\qquad{}\times\biggl[1\times\exp\bigl(-\lambda\bigl(u\mid
\mathbf{x}_i^h(u)\bigr)\bigr) + \biggl( \frac{\exp (-\lambda({u}\mid\mathbf
{x}_i^h({u})) )}{
1-\exp(-\lambda({j}\mid\mathbf{x}^h_i({u})) )}
\biggr)^2
\\
&&\quad\qquad{}\times \bigl(1-\exp\bigl(-\lambda\bigl(u\mid
\mathbf{x}_i^h(u)\bigr)\bigr) \bigr) \biggr]\nonumber
\\
\nonumber
&&\qquad = \sum_{u} \sum
_{i} Y_i(u-1)^2 \frac{\partial\lambda({u}\mid\mathbf{x}_i^h({u}))}{
\partial\theta_k}
\frac{\partial\lambda({u}\mid\mathbf{x}_i^h({u}))}{\partial{\theta_m}}
\\
&&\quad\qquad{}
\times\biggl( \frac{\exp (-\lambda({u}\mid\mathbf{x}_i^h({u})) )}{
1-\exp(-\lambda({j}\mid\mathbf{x}^h_i({u})) )} \biggr).\nonumber
\end{eqnarray}
In an analogous way, we can prove that the second element of equation
\eqref{eqFi,u,k} equals zero.
The third element of equation \eqref{eqFi,u,k} is the sum of two zero
terms as follows:
\begin{eqnarray}
\label{eqLHS2} && \mathbb{E} \biggl[ \sum_{u} \sum
_{u' \neq u} \sum_{i} F(i,u,k)
\times F\bigl(i,u',m\bigr) \biggr]\nonumber
\\
&&\qquad = \mathbb{E} \biggl[ \sum_{u}\sum
_{u' <u} \sum_{i} F(i,u,k)
\times F\bigl(i,u',m\bigr) \biggr] \nonumber
\\
&&\quad\qquad{} + \mathbb{E} \biggl[ \sum
_{u}\sum_{ u'>u} \sum
_{i} F(i,u,k) \times F\bigl(i,u',m
\bigr) \biggr]
\\
\nonumber
&&\qquad = \biggl[ \sum_{ u'<u} \sum
_{u} \sum_{i} F
\bigl(i,u',m\bigr) \times\underbrace{ \mathbb{E}\bigl[ F(i,u,k)
\bigr]}_{0} \biggr]
\\
&&\quad\qquad{}+ \biggl[ \sum_{u < u'}
\sum_{u} \sum_{i}
F(i,u,k) \times\underbrace{\mathbb{E}\bigl[ F\bigl(i,u',m\bigr)
\bigr]}_{0} \biggr] =0,\nonumber
\end{eqnarray}
where we used equation \eqref{eqproofC41}. In an analogous way, we can
prove that the last element of equation \eqref{eqFi,u,k} equals zero.
Now, from equations \eqref{eqFi,u,k}, \eqref{eqLHS1} and \eqref
{eqLHS2}, we can conclude that
\[
\mathbb{E} \biggl[ \sum_{u} \sum
_{u'} \sum_{i} \sum
_{i'} F(i,u,k) \times F\bigl(i',u',m
\bigr) \biggr]=\mathbb{E} \biggl[ \sum_{u} \sum
_{i} F(i,u,k) \times F(i,u,m) \biggr].
\]
We have finished simplifying the left-hand side of equation \eqref{eqc42}.
The right-hand side of equation (\ref{eqc42}), which equals the
expected value of the negative of the Hessian matrix of $\log\mathcal
{L}_N(\bolds{\theta}\mid X)$, can be rewritten as follows:
%
\begin{eqnarray}
\label{eqRHS}
&& \tiny \mathbb{E} \biggl[ - \frac{\partial^2}{\partial
\theta_k\,\partial\theta_m} \log
\mathcal{L}_N(\bolds{\theta}\mid X) \biggr]\nonumber
\\
&&\qquad = \sum_{u} \sum
_{i} \mathbb{E} \biggl[ Y_i(u-1)
\biggl((1-2y_{i,u}) \frac{\exp(-\lambda({u}\mid\mathbf{x}_i^h({u})) )}{
y_{i,u}+(1-2y_{i,u})\exp(-\lambda({u}\mid\mathbf{x}_i^h({u})) )} \biggr)\hspace*{-10pt}
\\
\nonumber
&&\quad\qquad {}\times\biggl( \frac{\partial^2\lambda({u}\mid\mathbf
{x}_i^h({u}))}{\partial\theta_k \,\partial\theta_m} - \frac{\frac
{\partial\lambda({u}\mid\mathbf{x}_i^h({u}))}{\partial\theta_k} \times
\frac{\partial\lambda({u}\mid\mathbf{x}_i^h({u}))}{\partial\theta_m}
\times y_{i,u}}{y_{i,u}+(1-2y_{i,u})\exp(-\lambda({u}\mid\mathbf
{x}_i^h({u})) )} \biggr)
\biggr].
\end{eqnarray}
Pulling the expectation into the sum and separating cases where
$y_{i,u}=1$ from $y_{i,u}=0$, we can simplify equation \eqref{eqRHS} as
%
\begin{eqnarray}
\label{eqb12} &&\mathbb{E} \biggl[ - \frac{\partial^2}{\partial\theta
_k\,\partial\theta_m} \log
\mathcal{L}_N(\bolds{\theta}\mid X) \biggr]\nonumber
\\
\nonumber
&&\qquad =\sum_{u} \sum
_{i} Y_i(u-1) \biggl[ \biggl( \frac{\partial^2\lambda({u}\mid\mathbf
{x}_i^h({u}))}{\partial\theta_k \,\partial\theta_m}
\biggr) \exp\bigl(-\lambda\bigl({u}\mid\mathbf{x}_i^h({u})
\bigr) \bigr)
\\
&&\quad\qquad{} - \frac{\exp(-\lambda({u}\mid\mathbf{x}_i^h({u})) )}{1-\exp
(-\lambda({u}\mid\mathbf{x}_i^h({u})) )}
\\
\nonumber
&&\quad\qquad{} \times\biggl( \frac{\partial^2\lambda({u}\mid\mathbf
{x}_i^h({u}))}{\partial\theta_k \,\partial\theta_m} - \frac{\frac
{\partial\lambda({u}\mid\mathbf{x}_i^h({u}))}{\partial\theta_k} \times
\frac{\partial\lambda({u}\mid\mathbf{x}_i^h({u}))}{\partial\theta_m}
}{1-\exp(-\lambda({u}\mid\mathbf{x}_i^h({u})) )} \biggr)
\bigl(1-\exp\bigl(-\lambda\bigl({u}\mid\mathbf{x}_i^h({u})
\bigr)\bigr) \bigr) \biggr]
\\
\nonumber
&&\qquad =\sum_{u} \sum
_{i} Y_i(u-1) \frac{\partial\lambda({u}\mid\mathbf{x}_i^h({u}))}{
\partial\theta_k}
\frac{\partial\lambda({u}\mid\mathbf{x}_i^h({u}))}{\partial_m} \times
\biggl( \frac{\exp (-\lambda({u}\mid\mathbf{x}_i^h({u})) )}{
1-\exp(-\lambda({j}\mid\mathbf{x}^h_i({u})) )} \biggr).
\end{eqnarray}
Since the last term in equation \eqref{eqb12} is equivalent to the last
term in equation \eqref{eqLHS1}, and equations \eqref{eqb12} and \eqref
{eqLHS1} are, respectively, the right- and left-hand sides of equation
\eqref{eqc42}, we can conclude that equation \eqref{eqc42} is
satisfied.

\item[(C5)] This condition is satisfied since we proved earlier that
the log-likelihood is strictly convex and, therefore, $\mathbf{I}(\bolds
{\theta})$ (which is the negative of the Hessian matrix) is positive
definite and $I_{k,m}(\bolds{\theta})$ is finite for all $k$ and $m$.

\item[(C6)] To show that {(C6)} holds true, we need to show that the
third derivatives are absolutely bounded.
We first show that there exits a positive real number $\mathcal{M}$
such that
\begin{eqnarray}
\biggl\llvert\frac{\partial{{\partial}^3}}{\partial{\theta_k}\,
\partial{\theta_m} \,\partial{\theta_z}} \log\mathcal{L}_N(\bolds{\theta
}\mid
X)\biggr\rrvert\leq\sum_i \sum
_{j} \bigl\llvert x_{i,k}(j)\bigr\rrvert\bigl
\llvert x_{i,m}(j)\bigr\rrvert\bigl\llvert x_{i,z}(j)\bigr
\rrvert\mathcal{M}\nonumber
\\
\eqntext{\mbox{for all } \bolds{\theta} \in\omega.}
\end{eqnarray}
Since $\bolds{\alpha},\bolds{\beta}$ and $X$ are all bounded, we can assume
that there exists a positive real number $\mathcal{M}_1$ such that
$\exp(\bolds{\alpha}^\top X)<\mathcal{M}_1$ and $\exp(\bolds{\beta
}^\top X)<\mathcal{M}_1$ for all $X$.
As a result, we can conclude using equations \eqref{eqthirdg1}--\eqref
{eqthirdg2} that
\[
\frac{\partial^3\lambda({j}\mid\mathbf{x}_i^h({j}))}{\partial\theta
_k\,\partial\theta_m \,\partial\theta_z} \leq\sum_{l=1}^{j}
\bigl\llvert x_{i,k}(l)\bigr\rrvert\bigl\llvert x_{i,m}(l)\bigr
\rrvert\bigl\llvert x_{i,z}(l)\bigr\rrvert\mathcal{M}_1.
\]
Also, as
$\frac{\partial{{\partial}^3}\log\mathcal{L}_N(\bolds{\theta}\mid
X)}{\partial{\theta_k} \,\partial{\theta_m} \,\partial{\theta_z}}$ is a
linear combination of some nonlinear functions all including
$x_{i,k}(j)$, $x_{i,m}(j)$ and $x_{i,z}(j)$, and as the sum and the
product of bounded functions are also bounded, there exits a real
number $\mathcal{M}$ for which
\[
\biggl\llvert\frac{\partial{\partial^3}}{\partial{\theta_k}
\,\partial{\theta_m} \,\partial{\theta_z}} \log\mathcal{L}_N(\bolds{\theta
}\mid
X)\biggr\rrvert\leq\sum_i \sum
_{j} \bigl\llvert x_{i,k}(j)\bigr\rrvert\bigl
\llvert x_{i,m}(j)\bigr\rrvert\bigl\llvert x_{i,z}(j)\bigr
\rrvert\mathcal{M}\qquad\forall\bolds{\theta} \in\omega.
\]
As feature values are all bounded, then
\begin{eqnarray*}
\mathbb{E} \biggl[\sum_i \sum
_{j} \bigl\llvert x_{i,k}(j)\bigr\rrvert\bigl
\llvert x_{i,m}(j)\bigr\rrvert\bigl\llvert x_{i,z}(j)\bigr
\rrvert \mathcal{M} \biggr]
& =& \sum_i \sum
_{j} \mathbb{E} \bigl\llvert x_{i,k}(j)
x_{i,m}(j) x_{i,z}(j) \bigr\rrvert \mathcal{M}
\\
&<& \infty,
\end{eqnarray*}
which can be used to prove that (C6) is satisfied. These
theorems together provide an immediate proof for Theorem \ref{theoremconsis}.\quad\qed
\end{longlist}\noqed
\end{pf}

\section{Table of results for simulation study}\label{app3}\vspace*{-10pt}
%
\begin{table}[h]
\tabcolsep=0pt
\caption{Table of results for simulation study---[mean, standard deviation (sd) and mean squared error (MSE)] for 5 values of $N$} \label{tablesim}
{\fontsize{8.5}{10.5}{\selectfont
\begin{tabular*}{\tablewidth}{@{\extracolsep{\fill}}@{}lcccc@{}}
\hline
\multicolumn{1}{@{}l}{\textbf{Parameters}} & $\bolds{\beta_0}$ & $\bolds{\alpha_0}$ & $\bolds{\beta_1}$ & $\bolds{\alpha_1}$\\
\hline
\multicolumn{1}{@{}l}{\textbf{True values}} & $\bolds{-7}$ & $\bolds{-14}$ & \textbf{0.5} & \textbf{5} \\
\hline
 \multicolumn{1}{@{}l}{$\bolds{N}$} & \multicolumn{4}{c}{\textbf{(Mean, sd, MSE)}}\\
\hline
 \phantom{0}50 & $(-7.19, 0.39, 0.19)$ & $(-15.17, 4.68, 23.27)$ & $(0.46, 0.69, 0.48)$ & $(5.39, 1.75, 3.22)$\\
 100 & $(-7.13, 0.26, 0.08)$ & $(-14.45, 2.52, 6.54)$ & $(0.45, 0.59, 0.36)$ & $(5.15, 0.87, 0.77)$\\
200 & $(-7.05, 0.18, 0.03)$ & $(-14.54, 1.53, 2.64)$ & $(0.49, 0.36, 0.13)$ & $(5.19, 0.54, 0.32)$ \\
400 & $(-7.00, 0.12, 0.02)$ & $(-14.26, 0.94, 0.95)$ & $(0.44, 0.27, 0.08)$ & $(5.09, 0.32, 0.11)$\\
 800 & $(-7.01, 0.09, 0.01)$ & $(-14.05, 0.66, 0.44)$ & $(0.49, 0.19, 0.04)$ & $(5.02, 0.23, 0.05)$ \\
\hline
\end{tabular*}\vspace*{-20pt}}}
\end{table}

\section{Input parameters of HMM used in Section~\texorpdfstring{\protect\ref{22222}}{6.2.3}}\label{app4}
\begin{eqnarray}
\mathbf{P} =
\lleft[\matrix{
0.9 &0.09 &0.01 \cr
0 &0.87& 0.13 \cr
0 & 0 & 1} \rright],\qquad
\mathbf{Q}  =
\lleft[\matrix{
0.6 &0.3 &0.05& 0.05& 0 \cr
0.1 &0.2& 0.4 &0.2& 0.1 \cr
0 &0.05 &0.05& 0.3 &0.6} \rright],\nonumber
\\
\eqntext{\zeta=20, \eta=4.5, \mbox{ and }\gamma=1.4.}
\end{eqnarray}
\end{appendix}

\section*{Acknowledgments}
We would like to thank the Accenture---MIT Alliance for funding this
work and ENEL for providing the data. We would specifically like to
thank Giuseppe Panunzio, Cristian Corbetti, Andrew Fano and Thania
Villatoro from Accenture. We would also like to thank \c{S}eyda Ertekin
and Ken Cohn for helpful discussions.

\begin{supplement}[id=suppA]
\stitle{Supplementary material for ``The latent state hazard model, with application to~wind turbine reliability''}
\slink[doi]{10.1214/15-AOAS859SUPP} 
\sdatatype{.pdf}
\sfilename{aoas859\_supp.pdf}
\sdescription{The supplementary material includes three sections:
A:~Interpretation of the model;
B:~Notes on the relativity assumption, and Supplement; and
C:~Making the coefficients interpretable.}
\end{supplement}

%

\printaddresses

\begin{thebibliography}{48}

\bibitem[\protect\citeauthoryear{Andersen and Vaeth}{1989}]{Anderson1989}
%
\begin{barticle}[author]
\bauthor{\bsnm{Andersen},~\bfnm{Per~Kragh}\binits{P.~K.}} \AND
\bauthor{\bsnm{Vaeth},~\bfnm{Michael}\binits{M.}}
(\byear{1989}).
\btitle{Simple parametric and nonparametric models for excess and
relative mortality}.
\bjournal{Biometrics}
\bvolume{45}
\bpages{523--535}.
\end{barticle}
%

\bptok{imsref}%
\endbibitem

\bibitem[\protect\citeauthoryear{Banjevic and Jardine}{2006}]{Banjevic2006}
%
\begin{barticle}[mr]
\bauthor{\bsnm{Banjevic},~\bfnm{D.}\binits{D.}} \AND
\bauthor{\bsnm{Jardine},~\bfnm{A.~K.~S.}\binits{A.~K.~S.}}
(\byear{2006}).
\btitle{Calculation of reliability function and remaining useful life
for a {M}arkov failure time process}.
\bjournal{IMA J. Manag. Math.}
\bvolume{17}
\bpages{115--130}.
\bid{doi={10.1093/imaman/dpi029}, issn={1471-678X}, mr={2216398}}
\end{barticle}
%

\bptok{imsref}%
\endbibitem

\bibitem[\protect\citeauthoryear{Banjevic et~al.}{2001}]{Banjevic2001}
%
\begin{barticle}[author]
\bauthor{\bsnm{Banjevic},~\bfnm{D.}\binits{D.}},
\bauthor{\bsnm{Jardine},~\bfnm{A.~K.~S.}\binits{A.~K.~S.}},
\bauthor{\bsnm{Makis},~\bfnm{V.}\binits{V.}} \AND
\bauthor{\bsnm{Ennis},~\bfnm{M.}\binits{M.}}
(\byear{2001}).
\btitle{A control-limit policy and software for condition-based
maintenance optimization}.
\bjournal{INFOR}
\bvolume{39}
\bpages{32--50}.
\end{barticle}
%

\bptok{imsref}%
\endbibitem

\bibitem[\protect\citeauthoryear{Bian and Gebraeel}{2012}]{Bian2012974}
%
\begin{barticle}[author]
\bauthor{\bsnm{Bian},~\bfnm{L.}\binits{L.}} \AND
\bauthor{\bsnm{Gebraeel},~\bfnm{N.}\binits{N.}}
(\byear{2012}).
\btitle{Computing and updating the first-passage time distribution for
randomly evolving degradation signals}.
\bjournal{IIE Transactions (Institute of Industrial Engineers)}
\bvolume{44}
\bpages{974--987}.
\end{barticle}
%

\bptok{imsref}%
\endbibitem

\bibitem[\protect\citeauthoryear{Bian and Gebraeel}{2013}]{Bian2013}
%
\begin{barticle}[mr]
\bauthor{\bsnm{Bian},~\bfnm{Linkan}\binits{L.}} \AND
\bauthor{\bsnm{Gebraeel},~\bfnm{Nagi}\binits{N.}}
(\byear{2013}).
\btitle{Stochastic methodology for prognostics under continuously
varying environmental profiles}.
\bjournal{Stat. Anal. Data Min.}
\bvolume{6}
\bpages{260--270}.
\bid{doi={10.1002/sam.11154}, issn={1932-1864}, mr={3062269}}
\end{barticle}
%

\bptok{imsref}%
\endbibitem

\bibitem[\protect\citeauthoryear{Boutros and Liang}{2011}]{Boutros20112102}
%
\begin{barticle}[author]
\bauthor{\bsnm{Boutros},~\bfnm{T.}\binits{T.}} \AND
\bauthor{\bsnm{Liang},~\bfnm{M.}\binits{M.}}
(\byear{2011}).
\btitle{Detection and diagnosis of bearing and cutting tool faults
using hidden Markov models}.
\bjournal{Mech. Syst. Signal Process.}
\bvolume{25}
\bpages{2102--2124}.
\end{barticle}
%

\bptok{imsref}%
\endbibitem

\bibitem[\protect\citeauthoryear{Collett}{2003}]{Collett2003}
%
\begin{bbook}[mr]
\bauthor{\bsnm{Collett},~\bfnm{David}\binits{D.}}
(\byear{2003}).
\btitle{Modelling Binary Data},
\bedition{2nd} ed.
\bseries{Chapman \& Hall/CRC Texts in Statistical Science Series}.
\bpublisher{Chapman \& Hall/CRC},
\blocation{Boca Raton, FL}.
\bid{mr={1999899}}
\end{bbook}
%

\bptok{imsref}%
\endbibitem

\bibitem[\protect\citeauthoryear{Cox}{1972}]{Cox1972}
%
\begin{barticle}[mr]
\bauthor{\bsnm{Cox},~\bfnm{D.~R.}\binits{D.~R.}}
(\byear{1972}).
\btitle{Regression models and life-tables}.
\bjournal{J. Roy. Statist. Soc. Ser. B}
\bvolume{34}
\bpages{187--220}.
\bid{issn={0035-9246}, mr={0341758}}
\bptnote{check related}%
\end{barticle}
%

\bptok{imsref}%
\endbibitem

\bibitem[\protect\citeauthoryear{DeCastro, Litt and Frederick}{2008}]{DeCastro2008}
%
\begin{binproceedings}[author]
\bauthor{\bsnm{DeCastro},~\bfnm{J.~A.}\binits{J.~A.}},
\bauthor{\bsnm{Litt},~\bfnm{J.~S.}\binits{J.~S.}} \AND
\bauthor{\bsnm{Frederick},~\bfnm{D.~K.}\binits{D.~K.}}
(\byear{2008}).
\btitle{A modular aero-propulsion system simulation of a large
commercial aircraft engine}.
In \bbooktitle{AIAA-2008-4579, 44th AIAA/ASME/SAE/ASEE Joint Propulsion
Conference \& Exhibit}. \blocation{Hartford, USA}.
\end{binproceedings}
%

\bptok{imsref}%
\endbibitem

\bibitem[\protect\citeauthoryear{Fisher and Lin}{1999}]{Fisher1999}
%
\begin{barticle}[pbm]
\bauthor{\bsnm{Fisher},~\bfnm{L.~D.}\binits{L.~D.}} \AND
\bauthor{\bsnm{Lin},~\bfnm{D.~Y.}\binits{D.~Y.}}
(\byear{1999}).
\btitle{Time-dependent covariates in the Cox proportional-hazards
regression model}.
\bjournal{Annu. Rev. Public Health}
\bvolume{20}
\bpages{145--157}.
\bid{doi={10.1146/annurev.publhealth.20.1.145}, issn={0163-7525},
pmid={10352854}}
\end{barticle}
%

\bptok{imsref}%
\endbibitem

\bibitem[\protect\citeauthoryear{Flory, Kharoufeh and
Gebraeel}{2014}]{Flory20141227}
%
\begin{barticle}[author]
\bauthor{\bsnm{Flory},~\bfnm{J.~A.}\binits{J.~A.}},
\bauthor{\bsnm{Kharoufeh},~\bfnm{J.~P.}\binits{J.~P.}} \AND
\bauthor{\bsnm{Gebraeel},~\bfnm{N.~Z.}\binits{N.~Z.}}
(\byear{2014}).
\btitle{A switching diffusion model for lifetime estimation in randomly
varying environments}.
\bjournal{IIE Transactions (Institute of Industrial Engineers)}
\bvolume{46}
\bpages{1227--1241}.
\end{barticle}
%

\bptok{imsref}%
\endbibitem

\bibitem[\protect\citeauthoryear{Gebraeel and Pan}{2008}]{Gebraeel2008539}
%
\begin{barticle}[author]
\bauthor{\bsnm{Gebraeel},~\bfnm{N.}\binits{N.}} \AND
\bauthor{\bsnm{Pan},~\bfnm{J.}\binits{J.}}
(\byear{2008}).
\btitle{Prognostic degradation models for computing and updating
residual life distributions in a time-varying environment}.
\bjournal{IEEE Trans. Reliab.}
\bvolume{57}
\bpages{539--550}.
\end{barticle}
%

\bptok{imsref}%
\endbibitem

\bibitem[\protect\citeauthoryear{Ghasemi, Yacout and
Ouali}{2007}]{Ghasemi2007989}
%
\begin{barticle}[author]
\bauthor{\bsnm{Ghasemi},~\bfnm{A.}\binits{A.}},
\bauthor{\bsnm{Yacout},~\bfnm{S.}\binits{S.}} \AND
\bauthor{\bsnm{Ouali},~\bfnm{M.~S.}\binits{M.~S.}}
(\byear{2007}).
\btitle{Optimal condition based maintenance with imperfect information
and the proportional hazards model}.
\bjournal{Int. J. Prod. Res.}
\bvolume{45}
\bpages{989--1012}.
\end{barticle}
%

\bptok{imsref}%
\endbibitem

\bibitem[\protect\citeauthoryear{Ghasemi, Yacout and
Ouali}{2010}]{Ghasemi201045}
%
\begin{barticle}[author]
\bauthor{\bsnm{Ghasemi},~\bfnm{A.}\binits{A.}},
\bauthor{\bsnm{Yacout},~\bfnm{S.}\binits{S.}} \AND
\bauthor{\bsnm{Ouali},~\bfnm{M.~S.}\binits{M.~S.}}
(\byear{2010}).
\btitle{Evaluating the reliability function and the mean residual life
for equipment with unobservable states}.
\bjournal{IEEE Trans. Reliab.}
\bvolume{59}
\bpages{45--54}.
\end{barticle}
%

\bptok{imsref}%
\endbibitem

\bibitem[\protect\citeauthoryear{Gorjian et~al.}{2009}]{Gorjian2009}
%
\begin{binproceedings}[author]
\bauthor{\bsnm{Gorjian},~\bfnm{N.}\binits{N.}},
\bauthor{\bsnm{Ma},~\bfnm{L.}\binits{L.}},
\bauthor{\bsnm{Mittinty},~\bfnm{M.}\binits{M.}},
\bauthor{\bsnm{Yarlagadda},~\bfnm{P.}\binits{P.}} \AND
\bauthor{\bsnm{Sun},~\bfnm{Y.}\binits{Y.}}
(\byear{2009}).
\btitle{A review on reliability models with covariates}.
In \bbooktitle{4th World Congress on Engineering Asset Management,
WCEAM 2009}
\bpages{385--397}.
\blocation{Athens, Greece}.
\end{binproceedings}
%

\bptok{imsref}%
\endbibitem

\bibitem[\protect\citeauthoryear{Guo et~al.}{2009}]{Guo2009}
%
\begin{barticle}[author]
\bauthor{\bsnm{Guo},~\bfnm{H.}\binits{H.}},
\bauthor{\bsnm{Watson},~\bfnm{S.}\binits{S.}},
\bauthor{\bsnm{Tavner},~\bfnm{P.}\binits{P.}} \AND
\bauthor{\bsnm{Xiang},~\bfnm{J.}\binits{J.}}
(\byear{2009}).
\btitle{Reliability analysis for wind turbines with incomplete failure
data collected from after the date of initial installation}.
\bjournal{Reliab. Eng. Syst. Saf.}
\bvolume{94}
\bpages{1057--1063}.
\end{barticle}
%

\bptok{imsref}%
\endbibitem

\bibitem[\protect\citeauthoryear{Hameed et~al.}{2009}]{Hameed2009}
%
\begin{barticle}[author]
\bauthor{\bsnm{Hameed},~\bfnm{Z.}\binits{Z.}},
\bauthor{\bsnm{Hong},~\bfnm{Y.~S.}\binits{Y.~S.}},
\bauthor{\bsnm{Cho},~\bfnm{Y.~M.}\binits{Y.~M.}},
\bauthor{\bsnm{Ahn},~\bfnm{S.~H.}\binits{S.~H.}} \AND
\bauthor{\bsnm{Song},~\bfnm{C.~K.}\binits{C.~K.}}
(\byear{2009}).
\btitle{Condition monitoring and fault detection of wind turbines and
related algorithms: A review}.
\bjournal{Renew. Sustain. Energy Rev.}
\bvolume{13}
\bpages{1--39}.
\end{barticle}
%

\bptok{imsref}%
\endbibitem

\bibitem[\protect\citeauthoryear{H{\"o}hle}{2009}]{Hoehle2009}
%
\begin{barticle}[mr]
\bauthor{\bsnm{H{\"o}hle},~\bfnm{Michael}\binits{M.}}
(\byear{2009}).
\btitle{Additive-multiplicative regression models for spatio-temporal
epidemics}.
\bjournal{Biom. J.}
\bvolume{51}
\bpages{961--978}.
\bid{doi={10.1002/bimj.200900050}, issn={0323-3847}, mr={2744450}}
\end{barticle}
%

\bptok{imsref}%
\endbibitem

\bibitem[\protect\citeauthoryear{Hontelez, Burger and
Wijnmalen}{1996}]{Hontelez1996267}
%
\begin{barticle}[author]
\bauthor{\bsnm{Hontelez},~\bfnm{J.~A.~M.}\binits{J.~A.~M.}},
\bauthor{\bsnm{Burger},~\bfnm{H.~H.}\binits{H.~H.}} \AND
\bauthor{\bsnm{Wijnmalen},~\bfnm{D.~J.~D.}\binits{D.~J.~D.}}
(\byear{1996}).
\btitle{Optimum condition-based maintenance policies for deteriorating
systems with partial information}.
\bjournal{Reliab. Eng. Syst. Saf.}
\bvolume{51}
\bpages{267--274}.
\end{barticle}
%

\bptok{imsref}%
\endbibitem

\bibitem[\protect\citeauthoryear{Jardine, Anderson and Mann}{1987}]{Jardine1987}
%
\begin{barticle}[author]
\bauthor{\bsnm{Jardine},~\bfnm{A.~K.~S.}\binits{A.~K.~S.}},
\bauthor{\bsnm{Anderson},~\bfnm{P.~M.}\binits{P.~M.}} \AND
\bauthor{\bsnm{Mann},~\bfnm{D.~S.}\binits{D.~S.}}
(\byear{1987}).
\btitle{Application of the Weibull proportional hazards model to
aircraft and marine engine failure data}.
\bjournal{Qual. Reliab. Eng. Int.}
\bvolume{3}
\bpages{77--82}.
\bid{doi={10.1002/qre.4680030204}}
\end{barticle}
%

\bptok{imsref}%
\endbibitem

\bibitem[\protect\citeauthoryear{Kalbfleisch and
Prentice}{2002}]{Kalbfleisch2002}
%
\begin{bbook}[mr]
\bauthor{\bsnm{Kalbfleisch},~\bfnm{John~D.}\binits{J.~D.}} \AND
\bauthor{\bsnm{Prentice},~\bfnm{Ross~L.}\binits{R.~L.}}
(\byear{2002}).
\btitle{The Statistical Analysis of Failure Time Data},
\bedition{2nd} ed.
\bpublisher{Wiley},
\blocation{Hoboken, NJ}.
\bid{doi={10.1002/9781118032985}, mr={1924807}}
\end{bbook}
%

\bptok{imsref}%
\endbibitem

\bibitem[\protect\citeauthoryear{Kharoufeh}{2003}]{Kharoufeh2003237}
%
\begin{barticle}[mr]
\bauthor{\bsnm{Kharoufeh},~\bfnm{Jeffrey~P.}\binits{J.~P.}}
(\byear{2003}).
\btitle{Explicit results for wear processes in a {M}arkovian environment}.
\bjournal{Oper. Res. Lett.}
\bvolume{31}
\bpages{237--244}.
\bid{doi={10.1016/S0167-6377(02)00229-8}, issn={0167-6377}, mr={1967296}}
\end{barticle}
%

\bptok{imsref}%
\endbibitem

\bibitem[\protect\citeauthoryear{Kharoufeh and Cox}{2005}]{Kharoufeh2005533}
%
\begin{barticle}[author]
\bauthor{\bsnm{Kharoufeh},~\bfnm{J.~P.}\binits{J.~P.}} \AND
\bauthor{\bsnm{Cox},~\bfnm{S.~M.}\binits{S.~M.}}
(\byear{2005}).
\btitle{Stochastic models for degradation-based reliability}.
\bjournal{IIE Transactions (Institute of Industrial Engineers)}
\bvolume{37}
\bpages{533--542}.
\end{barticle}
%

\bptok{imsref}%
\endbibitem

\bibitem[\protect\citeauthoryear{Kharoufeh, Finkelstein and
Mixon}{2006}]{Kharoufeh2006303}
%
\begin{barticle}[mr]
\bauthor{\bsnm{Kharoufeh},~\bfnm{Jeffrey~P.}\binits{J.~P.}},
\bauthor{\bsnm{Finkelstein},~\bfnm{Daniel~E.}\binits{D.~E.}} \AND
\bauthor{\bsnm{Mixon},~\bfnm{Dustin~G.}\binits{D.~G.}}
(\byear{2006}).
\btitle{Availability of periodically inspected systems with {M}arkovian
wear and shocks}.
\bjournal{J. Appl. Probab.}
\bvolume{43}
\bpages{303--317}.
\bid{doi={10.1239/jap/1152413724}, issn={0021-9002}, mr={2248566}}
\end{barticle}
%

\bptok{imsref}%
\endbibitem

\bibitem[\protect\citeauthoryear{Kusiak, Zhang and Verma}{2013}]{Kusiak2013}
%
\begin{barticle}[author]
\bauthor{\bsnm{Kusiak},~\bfnm{A.}\binits{A.}},
\bauthor{\bsnm{Zhang},~\bfnm{Z.}\binits{Z.}} \AND
\bauthor{\bsnm{Verma},~\bfnm{A.}\binits{A.}}
(\byear{2013}).
\btitle{Prediction, operations, and condition monitoring in wind energy}.
\bjournal{Energy}
\bvolume{60}
\bpages{1--12}.
\bid{doi={10.1016/j.energy.2013.07.051}}
\end{barticle}
%

\bptok{imsref}%
\endbibitem

\bibitem[\protect\citeauthoryear{Lehmann and Casella}{1998}]{lehmann1998}
%
\begin{bbook}[mr]
\bauthor{\bsnm{Lehmann},~\bfnm{E.~L.}\binits{E.~L.}} \AND
\bauthor{\bsnm{Casella},~\bfnm{George}\binits{G.}}
(\byear{1998}).
\btitle{Theory of Point Estimation},
\bedition{2nd} ed.
\bpublisher{Springer},
\blocation{New York}.
\bid{mr={1639875}}
\end{bbook}
%

\bptok{imsref}%
\endbibitem

\bibitem[\protect\citeauthoryear{Lin and Ying}{1995}]{Lin1995}
%
\begin{barticle}[mr]
\bauthor{\bsnm{Lin},~\bfnm{D.~Y.}\binits{D.~Y.}} \AND
\bauthor{\bsnm{Ying},~\bfnm{Zhiliang}\binits{Z.}}
(\byear{1995}).
\btitle{Semiparametric analysis of general additive-multiplicative
hazard models for counting processes}.
\bjournal{Ann. Statist.}
\bvolume{23}
\bpages{1712--1734}.
\bid{doi={10.1214/aos/1176324320}, issn={0090-5364}, mr={1370304}}
\end{barticle}
%

\bptok{imsref}%
\endbibitem

\bibitem[\protect\citeauthoryear{Lu et~al.}{2009}]{Lu2009}
%
\begin{binproceedings}[author]
\bauthor{\bsnm{Lu},~\bfnm{B.}\binits{B.}},
\bauthor{\bsnm{Li},~\bfnm{Y.}\binits{Y.}},
\bauthor{\bsnm{Wu},~\bfnm{X.}\binits{X.}} \AND
\bauthor{\bsnm{Yang},~\bfnm{Z.}\binits{Z.}}
(\byear{2009}).
\btitle{A review of recent advances in wind turbine condition
monitoring and fault diagnosis}.
In \bbooktitle{2009 IEEE Power Electronics and Machines in Wind Applications}.
\blocation{Lincoln, NE}.
\end{binproceedings}
%

\bptok{imsref}%
\endbibitem

\bibitem[\protect\citeauthoryear{Makis and Jardine}{1991}]{MAKIS1991}
%
\begin{barticle}[author]
\bauthor{\bsnm{Makis},~\bfnm{V.}\binits{V.}} \AND
\bauthor{\bsnm{Jardine},~\bfnm{A.~K.~S.}\binits{A.~K.~S.}}
(\byear{1991}).
\btitle{Computation of optimal policies in replacement models}.
\bjournal{IMA J. Manag. Math.}
\bvolume{3}
\bpages{169--175}.
\bid{doi={10.1093/imaman/3.3.169}}
\end{barticle}
%

\bptok{imsref}%
\endbibitem

\bibitem[\protect\citeauthoryear{Marquez et~al.}{2012}]{Marquez2012}
%
\begin{barticle}[author]
\bauthor{\bsnm{Marquez},~\bfnm{Fausto~Pedro~Garca}\binits{F.~P.~G.}},
\bauthor{\bsnm{Tobias},~\bfnm{Andrew~Mark}\binits{A.~M.}},
\bauthor{\bsnm{Prez},~\bfnm{Jess~Mara~Pinar}\binits{J.~M.~P.}} \AND
\bauthor{\bsnm{Papaelias},~\bfnm{Mayorkinos}\binits{M.}}
(\byear{2012}).
\btitle{Condition monitoring of wind turbines: Techniques and methods}.
\bjournal{Renew. Energy}
\bvolume{46}
\bpages{169--178}.
\bid{doi={10.1016/j.renene.2012.03.003}}
\end{barticle}
%

\bptok{imsref}%
\endbibitem

\bibitem[\protect\citeauthoryear{Martinussen and
Scheike}{2002}]{Martinussen2002}
%
\begin{barticle}[mr]
\bauthor{\bsnm{Martinussen},~\bfnm{Torben}\binits{T.}} \AND
\bauthor{\bsnm{Scheike},~\bfnm{Thomas~H.}\binits{T.~H.}}
(\byear{2002}).
\btitle{A flexible additive multiplicative hazard model}.
\bjournal{Biometrika}
\bvolume{89}
\bpages{283--298}.
\bid{doi={10.1093/biomet/89.2.283}, issn={0006-3444}, mr={1913959}}
\end{barticle}
%

\bptok{imsref}%
\endbibitem

\bibitem[\protect\citeauthoryear{Marvuglia and
Messineo}{2012}]{Marvuglia2012574}
%
\begin{barticle}[author]
\bauthor{\bsnm{Marvuglia},~\bfnm{A.}\binits{A.}} \AND
\bauthor{\bsnm{Messineo},~\bfnm{A.}\binits{A.}}
(\byear{2012}).
\btitle{Monitoring of wind farms' power curves using machine learning
techniques}.
\bjournal{Appl. Energy}
\bvolume{98}
\bpages{574--583}.
\end{barticle}
%

\bptok{imsref}%
\endbibitem

\bibitem[\protect\citeauthoryear{Moghaddass and Rudin}{2015}]{Moghaddasssup}
%
\begin{bmisc}[author]
\bauthor{\bsnm{Moghaddass},~\bfnm{R.}\binits{R.}} \AND
\bauthor{\bsnm{Rudin},~\bfnm{C.}\binits{C.}}
(\byear{2015}).
\bhowpublished{Supplement to ``The latent state hazard model, with
application to~wind turbine reliability.''
DOI:\doiurl{10.1214/15-AOAS859SUPP}}.
\end{bmisc}
%

\bptok{imsref}%
\endbibitem

\bibitem[\protect\citeauthoryear{Moghaddass and Zuo}{2012}]{Moghaddass201294}
%
\begin{barticle}[author]
\bauthor{\bsnm{Moghaddass},~\bfnm{R.}\binits{R.}} \AND
\bauthor{\bsnm{Zuo},~\bfnm{M.~J.}\binits{M.~J.}}
(\byear{2012}).
\btitle{A parameter estimation method for a condition-monitored device
under multi-state deterioration}.
\bjournal{Reliab. Eng. Syst. Saf.}
\bvolume{106}
\bpages{94--103}.
\end{barticle}
%

\bptok{imsref}%
\endbibitem

\bibitem[\protect\citeauthoryear{Peng and Dong}{2011}]{Peng2011237}
%
\begin{barticle}[author]
\bauthor{\bsnm{Peng},~\bfnm{Y.}\binits{Y.}} \AND
\bauthor{\bsnm{Dong},~\bfnm{M.}\binits{M.}}
(\byear{2011}).
\btitle{A prognosis method using age-dependent hidden semi-Markov model
for equipment health prediction}.
\bjournal{Mech. Syst. Signal Process.}
\bvolume{25}
\bpages{237--252}.
\end{barticle}
%

\bptok{imsref}%
\endbibitem

\bibitem[\protect\citeauthoryear{Pijnenburg}{1991}]{Pijnenburg1991}
%
\begin{barticle}[author]
\bauthor{\bsnm{Pijnenburg},~\bfnm{M.}\binits{M.}}
(\byear{1991}).
\btitle{Additive hazards models in repairable systems reliability}.
\bjournal{Reliab. Eng. Syst. Saf.}
\bvolume{31}
\bpages{369--390}.
\bid{doi={10.1016/0951-8320(91)90078-L}}
\end{barticle}
%

\bptok{imsref}%
\endbibitem

\bibitem[\protect\citeauthoryear{Qian and Wu}{2014}]{Qian2014317}
%
\begin{barticle}[author]
\bauthor{\bsnm{Qian},~\bfnm{X.}\binits{X.}} \AND
\bauthor{\bsnm{Wu},~\bfnm{Y.}\binits{Y.}}
(\byear{2014}).
\btitle{Condition based maintenance optimization for the hydro
generating unit with dynamic economic dependence}.
\bjournal{International Journal of Control and Automation}
\bvolume{7}
\bpages{317--326}.
\end{barticle}
%

\bptok{imsref}%
\endbibitem

\bibitem[\protect\citeauthoryear{Qiu et~al.}{2012}]{Qiu2012}
%
\begin{barticle}[author]
\bauthor{\bsnm{Qiu},~\bfnm{Y.~N.}\binits{Y.~N.}},
\bauthor{\bsnm{Feng},~\bfnm{Y.~H.}\binits{Y.~H.}},
\bauthor{\bsnm{Tavner},~\bfnm{P.~J.}\binits{P.~J.}},
\bauthor{\bsnm{Richardson},~\bfnm{P.}\binits{P.}},
\bauthor{\bsnm{Erdos},~\bfnm{G.}\binits{G.}} \AND
\bauthor{\bsnm{Chen},~\bfnm{B.}\binits{B.}}
(\byear{2012}).
\btitle{Wind turbine SCADA alarm analysis for improving reliability.}
\bjournal{Wind Energy}
\bvolume{15}
\bpages{951--966}.
\end{barticle}
%

\bptok{imsref}%
\endbibitem

\bibitem[\protect\citeauthoryear{Rashid and Shifa}{2009}]{Mamunur2009}
%
\begin{barticle}[author]
\bauthor{\bsnm{Rashid},~\bfnm{M.}\binits{M.}} \AND
\bauthor{\bsnm{Shifa},~\bfnm{Naima}\binits{N.}}
(\byear{2009}).
\btitle{Consistency of the maximum likelihood estimator in logistic
regression model: A different approach}.
\bjournal{Journal of Statistics}
\bvolume{16}
\bpages{1--11}.
\end{barticle}
%

\bptok{imsref}%
\endbibitem

\bibitem[\protect\citeauthoryear{Rudin and Vahn}{2014}]{Rudin2014}
%
\begin{bmisc}[author]
\bauthor{\bsnm{Rudin},~\bfnm{Cynthia}\binits{C.}} \AND
\bauthor{\bsnm{Vahn},~\bfnm{Gah-Yi}\binits{G.-Y.}}
(\byear{2014}).
\bhowpublished{The big data newsvendor: Practical insights from machine
learning.
Working paper}.
\end{bmisc}
%

\bptok{imsref}%
\endbibitem

\bibitem[\protect\citeauthoryear{Saxena et~al.}{2008}]{saxena2008}
%
\begin{binproceedings}[author]
\bauthor{\bsnm{Saxena},~\bfnm{A.}\binits{A.}},
\bauthor{\bsnm{Goebel},~\bfnm{K.}\binits{K.}},
\bauthor{\bsnm{Simon},~\bfnm{D.}\binits{D.}} \AND
\bauthor{\bsnm{Eklund},~\bfnm{N.}\binits{N.}}
(\byear{2008}).
\btitle{Damage propagation modeling for aircraft engine run-to-failure
simulation}.
In \bbooktitle{2008 International Conference on Prognostics and Health
Management, PHM 2008}.
\blocation{Denver, USA}.
\end{binproceedings}
%

\bptok{imsref}%
\endbibitem

\bibitem[\protect\citeauthoryear{Si et~al.}{2011}]{Si20111}
%
\begin{barticle}[mr]
\bauthor{\bsnm{Si},~\bfnm{Xiao-Sheng}\binits{X.-S.}},
\bauthor{\bsnm{Wang},~\bfnm{Wenbin}\binits{W.}},
\bauthor{\bsnm{Hu},~\bfnm{Chang-Hua}\binits{C.-H.}} \AND
\bauthor{\bsnm{Zhou},~\bfnm{Dong-Hua}\binits{D.-H.}}
(\byear{2011}).
\btitle{Remaining useful life estimation---A~review on the statistical
data driven approaches}.
\bjournal{European J. Oper. Res.}
\bvolume{213}
\bpages{1--14}.
\bid{doi={10.1016/j.ejor.2010.11.018}, issn={0377-2217}, mr={2795805}}
\end{barticle}
%

\bptok{imsref}%
\endbibitem

\bibitem[\protect\citeauthoryear{Wu and Ryan}{2011}]{Wu2011}
%
\begin{barticle}[author]
\bauthor{\bsnm{Wu},~\bfnm{X.}\binits{X.}} \AND
\bauthor{\bsnm{Ryan},~\bfnm{S.~M.}\binits{S.~M.}}
(\byear{2011}).
\btitle{Optimal replacement in the proportional hazards model with
semi-Markovian covariate process and continuous monitoring}.
\bjournal{IEEE Trans. Reliab.}
\bvolume{60}
\bpages{580--589}.
\end{barticle}
%

\bptok{imsref}%
\endbibitem

\bibitem[\protect\citeauthoryear{Yang, Court and Jiang}{2013}]{Yang2013}
%
\begin{barticle}[author]
\bauthor{\bsnm{Yang},~\bfnm{W.}\binits{W.}},
\bauthor{\bsnm{Court},~\bfnm{R.}\binits{R.}} \AND
\bauthor{\bsnm{Jiang},~\bfnm{J.}\binits{J.}}
(\byear{2013}).
\btitle{Wind turbine condition monitoring by the approach of SCADA data
analysis}.
\bjournal{Renewable Energy}
\bvolume{53}
\bpages{365--376}.
\end{barticle}
%

\bptok{imsref}%
\endbibitem

\bibitem[\protect\citeauthoryear{Yang et~al.}{2014}]{Yang2014}
%
\begin{barticle}[author]
\bauthor{\bsnm{Yang},~\bfnm{Wenxian}\binits{W.}},
\bauthor{\bsnm{Tavner},~\bfnm{Peter~J.}\binits{P.~J.}},
\bauthor{\bsnm{Crabtree},~\bfnm{Christopher~J.}\binits{C.~J.}},
\bauthor{\bsnm{Feng},~\bfnm{Y.}\binits{Y.}} \AND
\bauthor{\bsnm{Qiu},~\bfnm{Y.}\binits{Y.}}
(\byear{2014}).
\btitle{Wind turbine condition monitoring: Technical and commercial challenges}.
\bjournal{Wind Energy}
\bvolume{17}
\bpages{673--693}.
\bid{doi={10.1002/we.1508}}
\end{barticle}
%

\bptok{imsref}%
\endbibitem

\bibitem[\protect\citeauthoryear{Zaher et~al.}{2009}]{Zaher2009}
%
\begin{barticle}[author]
\bauthor{\bsnm{Zaher},~\bfnm{A.}\binits{A.}},
\bauthor{\bsnm{McArthur},~\bfnm{S.~D.~J.}\binits{S.~D.~J.}},
\bauthor{\bsnm{Infield},~\bfnm{D.~G.}\binits{D.~G.}} \AND
\bauthor{\bsnm{Patel},~\bfnm{Y.}\binits{Y.}}
(\byear{2009}).
\btitle{Online wind turbine fault detection through automated SCADA
data analysis}.
\bjournal{Wind Energy}
\bvolume{12}
\bpages{574--593}.
\end{barticle}
%

\bptok{imsref}%
\endbibitem

\bibitem[\protect\citeauthoryear{Zhao et~al.}{2010}]{Zhao2010}
%
\begin{barticle}[author]
\bauthor{\bsnm{Zhao},~\bfnm{X.}\binits{X.}},
\bauthor{\bsnm{Fouladirad},~\bfnm{M.}\binits{M.}},
\bauthor{\bsnm{B{\'{e}}renguer},~\bfnm{C.}\binits{C.}} \AND
\bauthor{\bsnm{Bordes},~\bfnm{L.}\binits{L.}}
(\byear{2010}).
\btitle{Condition-based inspection/replacement policies for
non-monotone deteriorating systems with environmental covariates}.
\bjournal{Reliab. Eng. Syst. Saf.}
\bvolume{95}
\bpages{921--934}.
\end{barticle}
%

\bptok{imsref}%
\endbibitem

\bibitem[\protect\citeauthoryear{Zhou, Serban and Gebraeel}{2011}]{Zhou20111586}
%
\begin{barticle}[mr]
\bauthor{\bsnm{Zhou},~\bfnm{Rensheng~R.}\binits{R.~R.}},
\bauthor{\bsnm{Serban},~\bfnm{Nicoleta}\binits{N.}} \AND
\bauthor{\bsnm{Gebraeel},~\bfnm{Nagi}\binits{N.}}
(\byear{2011}).
\btitle{Degradation modeling applied to residual lifetime prediction
using functional data analysis}.
\bjournal{Ann. Appl. Stat.}
\bvolume{5}
\bpages{1586--1610}.
\bid{doi={10.1214/10-AOAS448}, issn={1932-6157}, mr={2849787}}
\end{barticle}
%

\bptok{imsref}%
\endbibitem
\end{thebibliography}
\end{document}